\documentclass[12pt]{article}

\pdfoutput=1

\usepackage[utf8]{inputenc}
\usepackage{fullpage}
\usepackage{hyperref}
\usepackage{graphicx}
\usepackage{amssymb,amsmath}
\usepackage{lipsum}
\usepackage[sort&compress,numbers]{natbib}
\usepackage{hyperref}
\hypersetup{hidelinks}
\usepackage{wrapfig}
\usepackage[usenames,dvipsnames,table]{xcolor}
\usepackage[total={6.5in,9in},top=1in,headsep=0.1in,headheight=1in]{geometry}
\usepackage{lineno}

\newcommand\snowmass{
\begin{center}
  \rule[-0.2in]{\hsize}{0.01in}\\
  \rule{\hsize}{0.01in}\\
  \vskip 0.1in
  Submitted to the Proceedings of the US Community Study\\ 
  on the Future of Particle Physics (Snowmass 2021)\\
  \rule{\hsize}{0.01in}\\
  \rule[+0.2in]{\hsize}{0.01in}\\[-2em]
\end{center}
}
 
\usepackage[firstpage=true]{background}
\backgroundsetup{contents={\parbox{6.5in}{\snowmass}}, scale=1,placement=top,opacity=1,color=black,position={3.25in,1.2in}}

\usepackage{fancyhdr}
\fancypagestyle{plain}{%
  \fancyhf{}%
  \fancyhead[C]{}
  \fancyfoot[C]{\thepage}
}

\fancypagestyle{empty}{%
  \fancyhf{}%
  \fancyhead[C]{{\it Dark matter Physics from Halo Measurements}}
  \fancyfoot[C]{\thepage}
}
\usepackage{xspace}
\usepackage{aas_macros}

\newcommand{\unit}[1]{\ensuremath{\mathrm{\,#1}}\xspace}

\newcommand{\GeV}{\unit{GeV}}

\newcommand{\Msun}{\unit{M_\odot}}

\newcommand{\msun}        {\Msun}

\title{Snowmass2021 Cosmic Frontier White Paper: \\
Dark Matter Physics from Halo Measurements}
\date{}

\usepackage{authblk}

\author[1]{Keith Bechtol}
\affil[1]{Physics Department, University of Wisconsin-Madison, Madison, WI 53706, USA}
\author[2,3,4]{Simon Birrer}
\affil[2]{Kavli Institute for Particle Astrophysics and Cosmology, Stanford University, Stanford, CA 94305, USA}
\affil[3]{Department of Physics, Stanford University, Stanford, CA 94305, USA}
\affil[4]{SLAC National Accelerator Laboratory, Menlo Park, CA 94025, USA}
\author[5]{Francis-Yan Cyr-Racine}
\affil[5]{Department of Physics and Astronomy, University of New Mexico, Albuquerque, NM 87106, USA}
\author[6]{Katelin Schutz}
\affil[6]{Department of Physics \& McGill Space Institute, McGill University, Montr\'{e}al, QC H3A 2T8, Canada}

\author[7,8]{Susmita Adhikari}
\affil[7]{Department of Physics, Indian Institute of Science Education and Research, Pashan, Pune 411008, India}
\affil[8]{Department of Astronomy and Astrophysics, University of Chicago, Chicago, IL 60637, USA}

\author[9]{Mustafa Amin}
\affil[9]{Department of Physics and Astronomy, Rice University, Houston, TX 77005, USA}

\author[7,10]{Arka Banerjee}
\affil[10]{Fermi National Accelerator Laboratory, Batavia, IL 60510, USA}

\author[11]{Simeon Bird}
\affil[11]{Department of Physics and Astronomy, University of California, Riverside, CA 92521, USA}

\author[12]{Nikita Blinov}
\affil[12]{Department of Physics and Astronomy, University of Victoria, Victoria, BC V8P 5C2, Canada}

\author[13]{Kimberly K.~Boddy} 
\affil[13]{Department of Physics, The University of Texas at Austin, Austin, TX 78712, USA}

\author[14]{Celine Boehm}
\affil[14]{Sydney Consortium for Particle Physics and Cosmology, School of Physics, The University of Sydney, NSW 2006, Australia}

\author[15,16]{Kevin Bundy}
\affil[15]{Department of Astronomy and Astrophysics, University of California, Santa Cruz, CA 95064, USA}
\affil[16]{University of California Observatories, University of California, Santa Cruz, CA 95064, USA} 

\author[17]{Malte Buschmann}
\affil[17]{Department of Physics, Princeton University, Princeton, NJ 08544, USA}

\author[18]{Sukanya Chakrabarti}
\affil[18]{School of Physics and Astronomy, Rochester Institute of Technology, Rochester, NY 14623, USA}

\author[19]{David Curtin}
\affil[19]{Department of Physics, University of Toronto, Toronto ON, M5S 3H4, Canada}

\author[20]{Liang Dai}
\affil[20]{Department of Physics, University of California, Berkeley, CA 94720, USA}

\author[8,10,21]{Alex Drlica-Wagner}
\affil[21]{Kavli Institute for Cosmological Physics, University of Chicago, Chicago, IL 60637, USA}

\author[22]{Cora Dvorkin}
\affil[22]{Department of Physics, Harvard University, Cambridge, MA 02138, USA}
  
\author[23]{Adrienne L. Erickcek}
\affil[23]{Department of Physics and Astronomy, University of North Carolina at Chapel Hill, Chapel Hill, NC, 27599, USA}

\author[24]{Daniel Gilman}
\affil[24]{Department of Astronomy and Astrophysics, University of Toronto, Toronto ON, M5S 3H4, CA}

\author[6]{Saniya Heeba}

\author[25]{Stacy Kim}
\affil[25]{Department of Physics, University of Surrey, Guildford, GU2 7XH, UK}

\author[26,27]{Vid Ir\v{s}i\v{c}}
\affil[26]{Kavli Institute for Cosmology, University of Cambridge, Cambridge CB3 0HA, UK}
\affil[27]{Cavendish Laboratory, University of Cambridge, Cambridge CB3 0HE, UK}

\author[15]{Alexie Leauthaud}

\author[28]{Mark Lovell}
\affil[28]{1Center for Astrophysics and Cosmology, Science Institute, University of Iceland, 107 Reykjavík, Iceland}

\author[29]{Zarija Luki\'c}
\affil[29]{Lawrence Berkeley National Laboratory, Berkeley, CA 94720, USA}

\author[30]{Yao-Yuan Mao} 
\affil[30]{Department of Physics and Astronomy, Rutgers, The State University of New Jersey, Piscataway, NJ 08854, USA}

\author[2,3]{Sidney Mau}

\author[31]{Andrea Mitridate}
\affil[31]{Walter Burke Institute for Theoretical Physics, California Institute of Technology, Pasadena, CA 91125, USA}

\author[32]{Philip Mocz}
\affil[32]{Lawrence Livermore National Laboratory, Livermore, CA 94550, USA}

\author[33]{Julian B.~Mu\~{n}oz}
\affil[33]{Harvard-Smithsonian Center for Astrophysics, Cambridge, MA 02138, USA}

\author[34,35]{Ethan~O.~Nadler}
\affil[34]{Carnegie Observatories, Pasadena, CA 91101, USA}
\affil[35]{Department of Physics $\&$ Astronomy, University of Southern California, Los Angeles, CA, 90007, USA}

\author[36]{Annika H. G. Peter}
\affil[36]{CCAPP, Department of Physics, Department of Astronomy, The Ohio State University, Columbus, OH 43210, USA}

\author[37]{Adrian Price-Whelan}
\affil[37]{Center for Computational Astrophysics, Flatiron Institute, New York, NY 10010, USA}

\author[38]{Andrew Robertson}
\affil[38]{Jet Propulsion Laboratory, California Institute of Technology, Pasadena, CA 91109, USA}

\author[39]{Nashwan Sabti}
\affil[39]{Department of Physics, King's College London, Strand, London WC2R 2LS, UK}

\author[40]{Neelima Sehgal}
\affil[40]{Physics and Astronomy Department, Stony Brook University, Stony Brook, NY 11794, USA}

\author[8,21]{Nora Shipp}

\author[34]{Joshua D. Simon}

\author[41]{Rajeev Singh} 
\affil[41]{Institute of Nuclear Physics Polish Academy of Sciences, PL-31-342 Krak\'ow, Poland}

\author[37,42]{Ken Van Tilburg}
\affil[42]{Center for Cosmology and Particle Physics, Department of Physics, New York University, New York, NY 10003, USA}

\author[2,3,4]{Risa H. Wechsler}

\author[43]{Axel Widmark}
\affil[43]{Dark Cosmology Centre, Niels Bohr Institute, University of Copenhagen, 2200 Copenhagen N, Denmark}

\author[11]{Hai-Bo Yu}

\begin{document}

\maketitle

\begin{abstract}
\noindent The non-linear process of cosmic structure formation produces gravitationally bound overdensities of dark matter known as halos.
The abundances, density profiles, ellipticities, and spins of these halos can be tied to the underlying fundamental particle physics that governs dark matter at microscopic scales.
Thus, macroscopic measurements of dark matter halos offer a unique opportunity to determine the underlying properties of dark matter across the vast landscape of dark matter theories. This white paper summarizes the ongoing rapid development of theoretical and experimental methods, as well as new opportunities, to use dark matter halo measurements as a pillar of dark matter physics. 

\end{abstract}

\tableofcontents

\section{Executive Summary}

The standard collisionless, cold dark matter (CDM) paradigm predicts that dark matter in the late-time Universe is clustered into gravitationally bound overdensities called halos.
Dark matter halos are distributed along a roughly power-law mass spectrum extending from galaxy clusters ($10^{15} M_\odot$) to ultra-faint dwarf galaxies ($10^8 M_\odot$).
Furthermore, CDM predicts that dark matter clusters into gravitationally bound halos that are not massive enough to host galaxies, with the mass spectrum of halos extending to sub-solar mass scales. While CDM provides an excellent parametric description of the Universe on large spatial scales, it does not provide any underlying mechanism for the production or properties of dark matter. Many theoretical models describing the particle physics of dark matter predict that the simple collisionless CDM model breaks down at small physical scales.  
This paper highlights the key science opportunities and observational milestones for exploring the fundamental nature of dark matter using astrophysical measurements of dark matter halos in the next decade.\\

\noindent {\bf Key Science Opportunities}
\begin{enumerate}
    \item \textbf{Sensitivity to a broad range of models:} To date, all positive experimental evidence for the existence and properties of dark matter come from astrophysical observations.
    Measurements of the abundance, density profiles, and spatial distribution of dark matter halos offer sensitivity to an enormous range of dark matter models, and are complementary to both terrestrial experiments and indirect searches for dark matter annihilation and decay products.
    \item \textbf{Rapidly advancing phenomenology:} Our understanding of how the fundamental properties of dark matter at a microscopic scale impact structure formation throughout cosmic history is rapidly advancing. There is enormous potential to further develop the detailed phenomenology for a broader range of dark matter models, and to explore new regions of theory space with new and archival data.
    \item \textbf{Connection between dark matter halos and the early Universe:} The seeds of cosmological structure formation were established in the earliest moments after the Big Bang. As we measure the distribution of dark matter across a broader range of physical scales, we simultaneously learn about the initial conditions of the Universe and probe periods of cosmic history that might be inaccessible by other means.
\end{enumerate}

\noindent {\bf Key Experimental Milestones}
\begin{enumerate}
    \item \textbf{Precision measurements of galaxy-scale dark matter halos:} Current and near-future facilities will provide a detailed mapping between luminous galaxies and their invisible dark matter halos across 13 billions years of cosmic history and 7 orders of magnitude in dark matter halo mass. Detailed measurements of halo abundances and density profiles will provide a stringent test of the CDM paradigm.
    \item \textbf{Population studies of completely dark halos:} Within the next decade, several observational techniques will become sensitive to dark matter halos at or below the minimum mass required to host stellar populations. 
    Such completely dark halos offer unique advantages to constrain the microphysical properties of dark matter because their evolution is less affected by baryonic physics. Furthermore, many theoretical models of dark matter predict conspicuous deviations from collisionless CDM in low-mass halos. 
    \item \textbf{Testing dark models that predict enhanced central densities in solar- and planetary-mass-scale halos:} A suite of innovative and ambitious observational techniques can be used to search for solar- and planetary-mass-scale halos via their subtle gravitational effects. The discovery of such low mass halos would extend our knowledge of the matter power spectrum by many orders of magnitude in physical scale and would transform our understanding of both dark matter properties and the early Universe.
\end{enumerate}


This white paper is organized as follows. In Section~\ref{sec:theory}, we enumerate several broad classes of DM behavior that can arise in different theories of DM. The models under discussion are not meant to be exhaustive, and we do not commit to particular theories at the level of a Lagrangian; rather, we attempt to ``coarse grain'' in theory space to capture the most essential DM properties that can observably affect DM halos, with such properties then mapping back on to specific models. Having established the diverse range of possible DM behaviors, we describe the discovery potential of DM halo and subhalo observations spanning nearly thirty orders of magnitude in halo mass and highlight the complementarity of these scales to different kinds of DM behavior. We group the DM halos and subhalos into the following categories: cluster scale, galactic scale, the scale below the threshold of galaxy formation, and sub-stellar scale. These correspond to Sections~\ref{sec:cluster} -- \ref{sec:subsolar}, respectively. Brief concluding remarks follow in Section~\ref{sec:conclusion}.

\section{Classes of Dark Matter Behavior}\label{sec:theory}

The Standard Model (SM) of particle physics has been a stunning success, with certain predictions of the theory having been validated out to many decimal places. However, the SM (as it is presently known) lacks a description of dark matter (DM), whose presence can be inferred via its gravitational influence in a wide variety of astrophysical environments. The astrophysical discovery of DM has inspired a concerted multi-disciplinary effort to characterize the microphysical properties of DM. To date, however, all attempts to conclusively pin down the nature of DM have yielded null results, indicating that any non-gravitational behavior (for instance, interactions with particles in the SM) must either be very weak or arise only under specific conditions that have not yet been explored. The growing number of null searches for DM in various experimental settings has inspired fresh  theoretical ideas over the past decade since the last Snowmass process. It is now more clear than ever that the landscape of potential DM candidates is unfathomably diverse; for instance, if the DM is comprised of a single constituent (e.g., one particle species), then the allowed range of DM particle mass spans roughly ninety orders of magnitude and is constrained only by astrophysical observations. Many plausible theories of DM are not so simple: there is a very real possibility that DM exists as part of an entire “dark sector,” with additional particles and forces and with a rich structure analogous to the SM.

The complexity of the space of plausible DM theories is difficult to quantify, but it is abundantly clear that a one-size-fits-all approach is not suitable in any empirical attempts to characterize the nature of DM. Halos and subhalos --- loosely defined for the purposes of this White Paper as gravitationally self-bound clumps of DM --- are the most promising systems for testing many broad classes of DM behavior. Halo observations can characterize the properties of dark matter using gravity, the only force through which DM is known to interact, since halos have large enough accumulations of DM to leave a gravitational imprint on the SM matter that we \emph{can} observe. Halos can also be incredibly diverse and can appear on different lengthscales, at different times in the history of the Universe, and in different ambient environments, providing a wide range of conditions for testing many different theoretical possibilities. 

Given everything that we have inferred about cosmological structure formation and DM halos from observation, we know that on sufficiently large scales greater than $\mathcal{O}({\rm Mpc})$ the DM can be described within the paradigm of collisionless, cold DM (CDM) where the DM has negligible interactions and negligible thermal velocity. However, many theories of DM predict deviations from this CDM paradigm that would be manifest in DM halos on a range of scales. We broadly classify these deviations into two categories: (1) \emph{ab initio} deviations, where early-Universe DM behavior modifies the initial conditions for structure formation and hence alters the formation of DM halos and subhalos and (2) \emph{in situ} deviations where a relatively high density of DM in the present-day Universe allows certain DM behaviors to become manifest. 

In the remainder of this section, we discuss several broad features of DM theories with signatures from \emph{ab initio} DM behavior, \emph{in situ} DM behavior, and \emph{hybrid} models that are affected by both \emph{ab initio} and in \emph{situ behaviors}. This section is not meant to be exhaustive but rather serves to capture the diversity of possible DM characteristics that can affect observations of DM halos.  

\subsection{\emph{Ab Initio} Impact on Structure Formation }

\begin{figure}[t!]
    \hspace{-8mm}
    \includegraphics[width=1.1\textwidth]{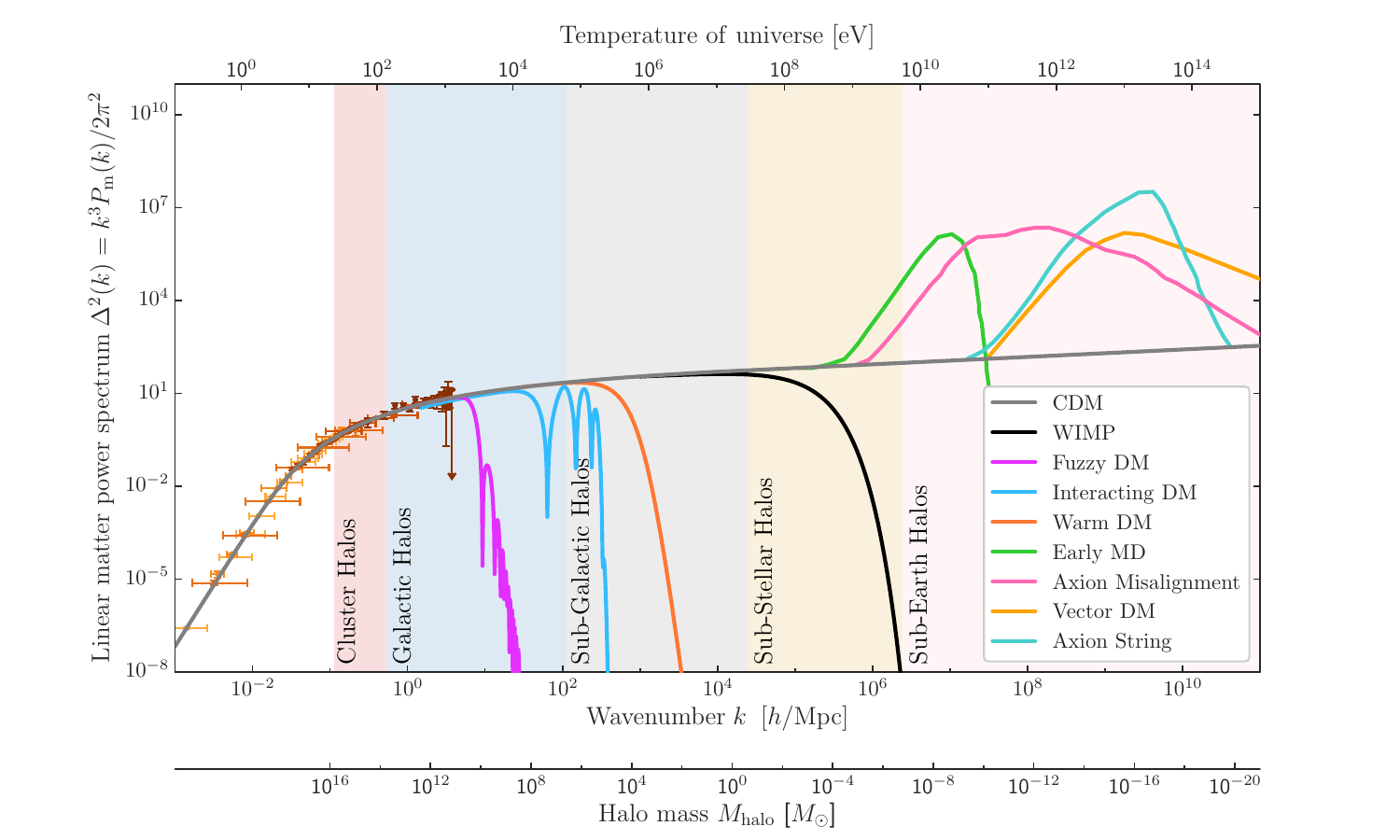}
    \caption{The dimensionless linear matter power spectrum extrapolated to $z=0$. Theoretical predictions are plotted for four models that suppress power: (1) ultra-light axion ``fuzzy'' dark matter with a mass $m=10^{-22}$~eV (magenta; \cite{Hu:2000}), (2) dark matter--baryon interactions with interaction cross section that scales with velocity as $ \sigma_0 v^{-4}$ for $\sigma_0 = 10^{-22}$cm$^2$ (blue; \cite{Boddy:2018kfv}), (3) thermal relic warm dark matter with a mass $m\sim 40$~keV (red; \cite{Viel:2005qj}), (4) weakly interacting massive particle dark matter represented by a bino-like neutralino with a mass $m\sim 100$~GeV (black; \cite{Green:2005fa}). Also shown are four models that enhance power on very small scales: (1) early matter domination assuming a reheat temperature of 10~MeV (green; \cite{Erickcek:2011us}), (2) post-inflationary production of QCD axions dominated by the misalignment mechanism (pink; \cite{Buschmann:2019icd}), (3) vector dark matter produced during inflation assuming an inflationary scale of $10^{14}$~GeV and a DM mass of $10^{-6}$~eV (orange; \cite{Graham:2015rva}), and (4) post-inflationary production of axions dominated by strings (cyan; \cite{Gorghetto:2020qws}). Note that the position of the power spectrum cutoff and/or enhancement depends on model parameters and is flexible for most cases shown here. Power spectrum measurements on large scales are compiled from \cite{Chabanier:2019eai}. Shaded vertical bands roughly indicate the characteristic kinds of halos formed on each scale, and the horizontal axes indicate wavenumber, halo mass, and the temperature of the Universe when that mode entered the horizon.}
    \label{fig:abinitio}
\end{figure}
\subsubsection{Thermal history}
\label{sec:WDM}

The cosmological clustering of dark matter on the scale of dark matter halos is set in large part by the DM thermal history in the early Universe. CDM predicts an abundance of halos down to arbitrarily small mass scales, but this will not be the case even for the simplest plausible particle physics models of DM. Thermal DM that becomes non-relativistic soon after its relic abundance is set (i.e. via decoupling from the SM plasma) is able to condense into low mass structures. Conversely, if the dark matter retains a relatively large kinetic energy, it will free-stream through low-mass density perturbations with low escape velocities and thus inhibit the formation of halos until the DM speed drops, either via kinetic decoupling or Hubble damping. 

If DM is made of weakly interacting massive particles (WIMPs) that are produced while non-relativistic via the freeze-out mechanism, then kinetic decoupling typically happens close to the freeze-out temperature for chemical decoupling, $T_f \sim m_\text{DM}/20$. Many supersymmetric theories naturally contain WIMP DM candidates, whose kinetic decoupling from the SM plasma would leave an imprint on structure formation on very small scales. For instance, for a bino-like neutralino with a mass around 100~GeV, the smallest halos that could form would be around an Earth mass~\cite{Green:2003un,2020Natur.585...39W}; structure on even smaller scales would be erased by the thermal motions of DM prior to decoupling. Thermal freeze-out production is excluded for DM below $\sim 10$~MeV based on observations of the CMB and from BBN~\cite{2020JCAP...01..004S,Boehm:2013jpa,Nollett:2013pwa,Green:2017ybv,Knapen:2017xzo,Giovanetti:2021izc,An:2022sva}, so exploring new freeze-out parameter space from structure formation will remain an enticing challenge for the future.

In a scenario where DM decouples from the SM while still relativistic, which is thermal history very similar to that of neutrinos, then the suppression of structure can occur on even larger scales. In this case, which is known as warm DM (WDM), structure can only form once the DM velocity is sufficiently Hubble damped, which suppresses structure on small scales~\cite{2001ApJ...556...93B,viel2005constraining,Schneider:2011yu}. Currently, these effects are being probed primarily on scales near the threshold of galaxy formation, as discussed in Sec.~\ref{sec:beyond_galaxy_threshold}. This spatial scale is too large to probe WIMP decoupling, however headway is being made in constraining new WDM parameter space above the keV mass scale and will continue over the next decade up to masses of $\sim20$~keV~\cite{Irsic:2017ixq,Yeche:2017upn,Baur:2017stq,Hsueh:2019ynk,Gilman:2019nap,Banik:2019cza,Banik:2019smi,Nadler:2019zrb,Nadler:2020prv,Rogers:2020cup,Pedersen:2020kaw,Enzi:2020ieg,Nadler:2021dft,Drlica-Wagner:2019xan,Munoz:2019hjh}. 

A final thermal history relevant for structure formation is DM freeze-in, where DM is produced non-thermally from SM particles in the thermal bath of the early Universe annihilating and decaying. A classic example of a DM candidate born via freeze-in is the sterile neutrino~\cite{1994PhRvL..72...17D,2008JCAP...06..031L}, but freeze-in is now known to be a more general mechanism that can be realized in a number of different models in the limit of small DM-SM couplings~\cite{Hall:2009bx,Chu:2011be,Bernal:2017kxu}. Freeze-in DM below the GeV mass scale is a key benchmark of the low-mass DM direct detection program, and can simultaneously be explored via clustering signatures. The phase-space distribution of freeze-in DM can vary substantially between models because the phase space is non-thermal, but current constraints and near-future projections are at the level of 10-100~keV~\cite{Dvorkin:2019zdi,Dvorkin:2020xga,Ballesteros:2020adh,DEramo:2020gpr,Baumholzer:2020hvx,Dienes:2021cxp,Decant:2021mhj}. Given the complementarity with upcoming direct detection experiments, the exploration of this parameter space is a timely goal for the coming decade.

\subsubsection{Non-thermal history}\label{sec:non-therm}

Besides the thermal history of dark matter, there are a range of non-thermal scenarios with intriguing signatures. One of such scenarios involves axion minihalos which occur naturally for any axion as long as the underlying Pecci--Quinn symmetry is broken after inflation. These axion minihalos are seeded purely by self-interactions and topological defects in the radiation era~\cite{Kibble_1976,PhysRevD.26.435,PhysRevLett.48.1867,PhysRevD.35.1138,DAVIS1986225,HARARI1987361,PhysRevD.32.3172,BATTYE1994260,PhysRevD.85.105020,PhysRevD.91.065014,Fleury_2016,Klaer_2017,Vaquero_2019,PhysRevLett.124.161103,Gorghetto:2018myk,10.21468/SciPostPhys.10.2.050,buschmann2021dark,PhysRevLett.124.021301,PhysRevD.103.103534,ohare2021simulations}, where overdensities in the axion field then collapse gravitationally at later times during the matter era~\cite{HOGAN1988228,PhysRevLett.71.3051,PhysRevD.49.5040,PhysRevD.75.043511,PhysRevD.50.769,Hardy:2016mns,PhysRevD.93.123509,Enander_2017}. The resulting halos are small, with a typical mass of $10^{-10}\,M_\odot$ and a Solar-System-sized radius. As the axion is a cold dark matter candidate, these minihalos form in addition to the much heavier thermally created halos. Despite their miniscule size, axion minihalos are likely to hinder direct detection experiments as a large amount of mass could be bound inside of them~\cite{Tinyakov_2016,Dokuchaev:2017psd,PhysRevD.95.063017,Knirck:2018knd,PhysRevD.104.063038}. However, several entirely new search strategies have been highlighted recently as well~\cite{Kolb_1996,Tkachev:2014dpa,Pshirkov:2016bjr,PhysRevD.97.083502,Katz_2018,Dai_2020,Croon:2020ouk,Croon:2020wpr,PhysRevLett.127.131103,Sun:2021oqp,Buen-Abad:2021qvj}. First proper estimates for the mass spectrum and abundance of axion minihalos have been obtained only recently. As this often requires large-scale numerical simulations due to non-linearities in the axion equation of motion, more computational work is needed in the future to confirm and refine these results. Furthermore, these halos are volatile against tidal disruptions within the Milky Way halo, though recent studies have found that a decent amount can survive until the present day~\cite{PhysRevLett.125.041301,PhysRevD.101.083014,PhysRevD.103.083525,Blinov:2019jqc,PhysRevD.104.023515}. More simulation work is needed on all fronts to properly understand axion minihalos and to confirm the feasibility of experimental search strategies.
 
 If the Pecci--Quinn symmetry is broken before inflation, it is still possible for various compact structures to emerge non-gravitationally should the initial misalignment angle be large~\cite{Arvanitaki:2019rax}. For axion masses above $10^{-5}$ eV those compact objects can be numerous enough to disrupt direct detection experiments. This does not only lead to unique new signatures but could also explain the non-observation of axions so far due to the potential lack of unbound mass. For smaller axion masses between $10^{-22}$ eV and $10^{-5}$ eV possible observable signatures include gravitional waves, disruptions of baryonic structures and an accelerated early Universe star formation. Numerical simulations will be required to quantitatively understand these signatures as the fraction of bound mass and the exact halo mass spectrum is not known.
 
 Inflationary particle production provides another compelling non-thermal mechanism for the enhancement of small-scale structure. This is a very minimal and generic way of producing the DM of the Universe since it only requires the gravitational coupling of a stable particle with mass less than the inflationary scale, $H_I$. Particle production occurs because the early-time vacuum state does not correspond to a vacuum state for an observer at late times in an expanding background.
  The resulting power spectrum of field fluctuations is strongly dependent on the spin of the particle; since the fluctuations of the DM field are uncorrelated with the inflaton, the DM density perturbations are isocurvature. 
  The latter fact is important because, for example, gravitational production of scalar particles $\phi$ leads to a scale-invariant dimensionless power spectrum, $\Delta_\phi^2 = (H_I/2\pi)^2$, which can conflict with CMB isocurvature constraints~\cite{Fox:2004kb,Planck:2018jri}. In contrast, the inflationary production of  a (longitudinal) vector boson, $A_L$, gives rise to a fluctuation spectrum $\Delta_{A_L}^2 = [k H_I/(2\pi m)]^2$ which is negligible at CMB scales (i.e., low-$k$), thereby automatically avoiding isocurvature constraints~\cite{Graham:2015rva}. Gravitational particle production also predicts the relic abundance as a function of the boson mass and the scale of inflation: $\Omega_{A_L}/\Omega_{cdm} \sim \sqrt{m/(10^{-5}\mathrm{eV})} [H_I /( 10^{14}\,\GeV)]^2$~\cite{Graham:2015rva,Ema:2019yrd,Ahmed:2020fhc,Kolb:2020fwh}; 
  CMB bounds on $H_I$ imply a lower bound $m\gtrsim 10^{-5}$ eV to saturate the relic abundance via this mechanism.\footnote{The calculation of the relic abundance is modified if post-inflationary reheating is non-instantaneous~\cite{Kolb:2020fwh}.} The scale-dependent field spectrum produces a density power spectrum that peaks at $k \sim a_* m$ with an $\mathcal{O}(1)$ magnitude, where $a_*$ is the scale factor when $H(a_*) = m$; the density power spectrum falls off as $k^3$ below the peak, and as $k^{-1}$ above. This leads to the gravitational formation of minihalos beginning near the time of matter-radiation equality (MRE). Note that because the power spectrum is enhanced for a wide range of scales (not just at $k \sim a_* m$), and the smallest minihalos formed at MRE are hierarchically assembled into larger and larger minihalos, resulting in a broad minihalo mass function~\cite{Blinov:2021axd}.

\subsubsection{Early departures from radiation domination}
\label{sec:non_standard_cosmology}

There is a troubling gap in the cosmological record: we do not know how the Universe evolved between the end of inflation and the onset of Big Bang nucleosynthesis (BBN) \cite{Allahverdi:2020bys}.  While it is customary to assume that the Universe became radiation dominated shortly after inflation ended and remained so for the next 50,000 years, there is no evidence than the expansion history of the early Universe was so straightforward.  As long as the Universe was radiation dominated at a temperature of $\sim\!\!5\,{\rm MeV}$, thermal neutrino production is sufficient to produce the observed abundances of light elements \citep{Hasegawa:2019jsa} and the observed CMB and matter power spectra \citep{deSalas:2015glj}.  Furthermore, deviations from radiation domination prior to BBN are strongly motivated.  After inflation ends, it is generally assumed that the inflaton oscillates and then decays, transferring its energy to a radiative plasma \cite{2010ARNPS..60...27A}.  While a scalar field oscillates around the minimum of a quadratic potential, it has zero pressure, so most inflationary scenarios include an early matter-dominated era (EMDE) between the end of inflation and the decay of the inflaton. After the inflaton decays into radiation, other oscillating scalar fields may come to dominate the energy density of the Universe; this phenomenon is a generic consequence of stabilized moduli in string theories \cite{2015IJMPD..2430022K}.  It is also possible that unstable massive particles could temporarily dominate the Universe prior to BBN, and hidden sector theories of dark matter frequently include EMDEs \cite{Zhang:2015era, Berlin:2016vnh,Berlin:2016gtr, Dror:2016rxc, Dror:2017gjq, Erickcek:2020wzd, Erickcek:2021fsu}.  Another possible deviation from radiation domination is a period of kination, during which the kinetic energy of a scalar field dominates the energy density of the Universe \cite{Spokoiny:1993kt, Joyce:1996cp, Ferreira:1997hj}.

Our ignorance of the pre-BBN expansion history profoundly limits our understanding of the origins of the baryon asymmetry \citep[e.g.][]{2001PhRvD..64b3508G, 2010PhRvD..82c5004A,2010PhRvL.105u1304D} and the origins of dark matter.  If dark matter chemically decouples during kination, $\langle \sigma v \rangle \gtrsim 3\times10^{-26}$ cm$^{3}$ s$^{-1}$ is required to generate the observed dark matter density \cite{Profumo:2003hq,Pallis:2005hm, DEramo:2017gpl, Redmond:2017tja}.
Conversely, entropy production during an EMDE dilutes the relic abundance of dark matter, so a smaller annihilation cross section is required to generate the observed dark matter density \citep{2001PhRvD..64b3508G}.  However, if the dark matter is also produced nonthermally (as a decay product, for instance), $\langle \sigma v \rangle$ may need to be increased so that any excess dark matter is eliminated \citep{2006PhRvD..74b3510G}. We are therefore faced with a dark degeneracy: we cannot determine the annihilation cross section required to produce the observed dark matter density without knowing the thermal history of the pre-BBN Universe.  A period of kination or an EMDE also relaxes constraints on axion \citep{2001PhRvD..64b3508G, 2008PhRvD..77h5020G, 2010PhRvD..81f3508V} and neutrino \citep{2001PhRvD..64b3508G, 2001PhRvD..64d3512G,2007JHEP...06..002Y} dark matter.  

Fortunately, sub-solar-mass dark matter microhalos could provide a unique window into the pre-BBN Universe that can be used to fill in this gap in the cosmological record.  During an EMDE or a period of kination, subhorizon perturbations in the dark matter density grow much faster than they would during radiation domination \cite{Erickcek:2011us, Redmond:2018xty}.  If $T_{\mathrm{RH}}$ is the radiation temperature when the Universe became radiation dominated, then an EMDE or a period of kination enhances the abundance of microhalos with masses less than $30 M_\oplus \,\left[{(10 \, \mbox{MeV)}}/{T_\mathrm{RH}}\right]^3$ \cite{Erickcek:2011us}.
The formation time and size of the smallest microhalos depends on the temperature of the dark matter particles \cite{Fan:2014zua, 2008JCAP...10..002G, Erickcek:2015bda, Waldstein:2016blt}, the properties of the particles that dominated the energy density during the EMDE \cite{Blanco:2019eij, Erickcek:2020wzd, Erickcek:2021fsu}, and the duration of the EMDE \cite{Blanco:2019eij,Barenboim:2021swl}. If dark matter is sufficiently cold and density perturbations remain in the linear regime during the EMDE, most of the dark matter is contained in microhalos at redshifts as high as 200, long before halos are expected to form in standard cosmologies \cite{Erickcek:2015jza, Erickcek:2015bda, Blanco:2019eij}.  
These early-forming microhalos are extremely dense, and they boost the dark matter annihilation rate by several orders of magnitude.  Consequently, the upper bounds on the dark matter annihilation rate established by gamma-ray observations can be used to constrain EMDEs \cite{Erickcek:2015jza,Erickcek:2015bda,Blanco:2019eij,StenDelos:2019xdk}.  In the future, EMDEs may also be constrained by pulsar timing arrays \cite{Lee:2020wfn, Delos:2021rqs} and observations of stellar microlensing within galaxy clusters \cite{Blinov:2021axd}, as detailed in Section~\ref{sec:subsolar}.

\subsubsection{Interactions with the SM}\label{subsec:bar_neut_scat}

While we only have gravitational evidence for dark matter so far, it is entirely plausible that dark matter interacts with SM particles. There have been numerous efforts to search for interactions between particle-like DM and the SM through collider experiments, direct detection, and indirect detection, but no concrete evidence has been found.
Cosmological and astrophysical observations offer a variety of avenues to test DM physics in regions of parameter space that both overlap with and are inaccessible to the more traditional searches.
One such avenue is to investigate the effects that interactions in the early Universe have on the subsequent formation of DM structure and substructure. In particular, DM elastic scattering with SM relativistic particles (i.e., photons and neutrinos) or nonrelativistic matter (generically referred to as ``baryons'') generically hinder the growth of structure and thus reduce DM halo abundances on corresponding scales.

The impact of DM interactions with ordinary particles is two-fold. First, collisions permit heat and momentum exchange between dark matter and the species it interacts with, thereby damping their primordial fluctuations. Second, they alter the free-streaming properties of DM and of the interacting species, either by inducing free streaming of DM particles (mixed damping~\cite{Boehm:2000gq}) or by delaying free streaming.
This second effect, important for light DM coupled to relativistic particles (i.e. photons and neutrinos), also affects the clumping of DM and of the interacting species.
The overall effect of collisional damping (analogous to Silk damping) usually translates into a reduction of power at progressively smaller scales, thereby leading to a smaller number of small-scale structures both in the Universe and within complex objects such as large galaxies.

The magnitude of the effect can be understood intuitively by an analytical approximation of the damping scale (at least in the simplest case, where the viscosity is dominant), namely:
$$ l_{cd}^2 \simeq \sum_i \int^{t_{dec(dm-i)}}  \frac{\rho_i v_i^2 }{a(t) \Gamma_i} $$ 
where $t_{dec(dm)}$ stands for the dark matter kinetic decoupling time with  species $i$, $\rho_i$ is the energy density of species $i$, $v_i$ its velocity and $\Gamma_i$ its interaction rate that is $\Gamma_i = \sum_j \Gamma_{ij}$ and $a(t)$ is the scale-factor. The larger the density $\rho_i$ is and the lighter the species are, the bigger the effect is --- unless the dark matter decouples from $i$ quickly. 
Since the DM decoupling is  given that the condition $\sigma v_{dm-i} \ n_i \simeq H$ (with $n_i$ the number density of species $i$ and $\sigma v_{dm-i}$ the elastic scattering cross section between the DM and species $i$), one can immediately see that the damping scale is the largest for the two relativistic species of the SM of Particle Physics, namely the neutrinos and photons which turn out to also have the largest densities $n_i$  and $\rho_i$,  unless $\sigma v_{dm-i}$ is small. In fact studies have shown that, in scenarios where the DM interacts with photons~\cite{Boehm:2000gq,Boehm:2001hm,Boehm:2004th,Wilkinson:2013kia,Stadler:2018jin} or neutrinos~\cite{Boehm:2000gq,Boehm:2004th,Mangano:2006mp,Wilkinson:2014ksa,Olivares-DelCampo:2017feq,Stadler:2019dii} with a relatively large cross section, the Universe may not contain any small galaxies and, in extreme scenarios where the cross section is too large, the Universe may not even contain enough large galaxies, thereby providing a means to exclude certain interaction strengths, otherwise invisible by more conventional means.  

Dark matter--baryon interactions produce a similar effect due to collisional damping~\cite{Dvorkin:2013cea,Xu:2018efh,Nadler:2019zrb,Nadler:2020prv,Maamari:2020aqz,Nguyen:2021cnb,Rogers:2021byl}.
Since the baryons are nonrelativistic and their density is relatively small, damping effects are negligible at large scales. At small scales, DM interactions produce a cutoff in the matter power spectrum, relative to $\Lambda$CDM.
The cutoff behavior is strikingly similar to that of warm dark matter~\cite{Nadler:2019zrb}, though there are subtleties related to the velocity dependence of the scattering cross section~\cite{Maamari:2020aqz}.
In particular, cross sections with a steep power-law velocity dependence recover substantial power after the initial cutoff, akin to dark acoustic oscillations.
A small-scale suppression of the matter power spectrum implies the existence of a minimum halo mass.
Constraints on the scattering cross section as a function of mass have been placed using current observations of Milky Way substructure~\cite{Nadler:2019zrb,Nadler:2020prv}.
Future observatories will probe even smaller substructures, extending experimental sensitivity to DM interactions~\cite{LSSTDarkMatterGroup:2019mwo}.

Once the dark matter interactions stopped, small-scale dark matter fluctuations can be further washed out by the dark matter free-streaming. Note that in most scenarios, the dark matter happens to be among the first species to kinetically decouple and thus the first to free-stream. However this is not necessarily true when the dark matter interacts with neutrinos and is relatively light (in the MeV range). There are in fact two interesting situations that are worth studying, namely: i) the neutrinos last interactions occur with the electrons, as it is expected in the Standard Model of Cosmology. In this case, the dark matter fluctuations are erased owing to the dark matter coupling to free-streaming neutrinos --- this is called the mixed damping effect in \cite{Boehm:2000gq}; ii) the neutrinos continue to interact with the dark matter after they stop interacting with electrons. In this case, the neutrinos  experience a longer period of collisional damping than in the SM and the start of their free-streaming period is delayed. Both situations lead to interesting effects. However the second case in particular can increase the number of large-scale-structures, which eventually has an implication on the estimates of the cosmological parameters (in particular  $N_{\rm eff}$ and $H_0$).

\subsection{\emph{In Situ} Impact on Structure Formation }
\label{insitu}
\begin{figure}
    \centering
    \includegraphics[width=\textwidth]{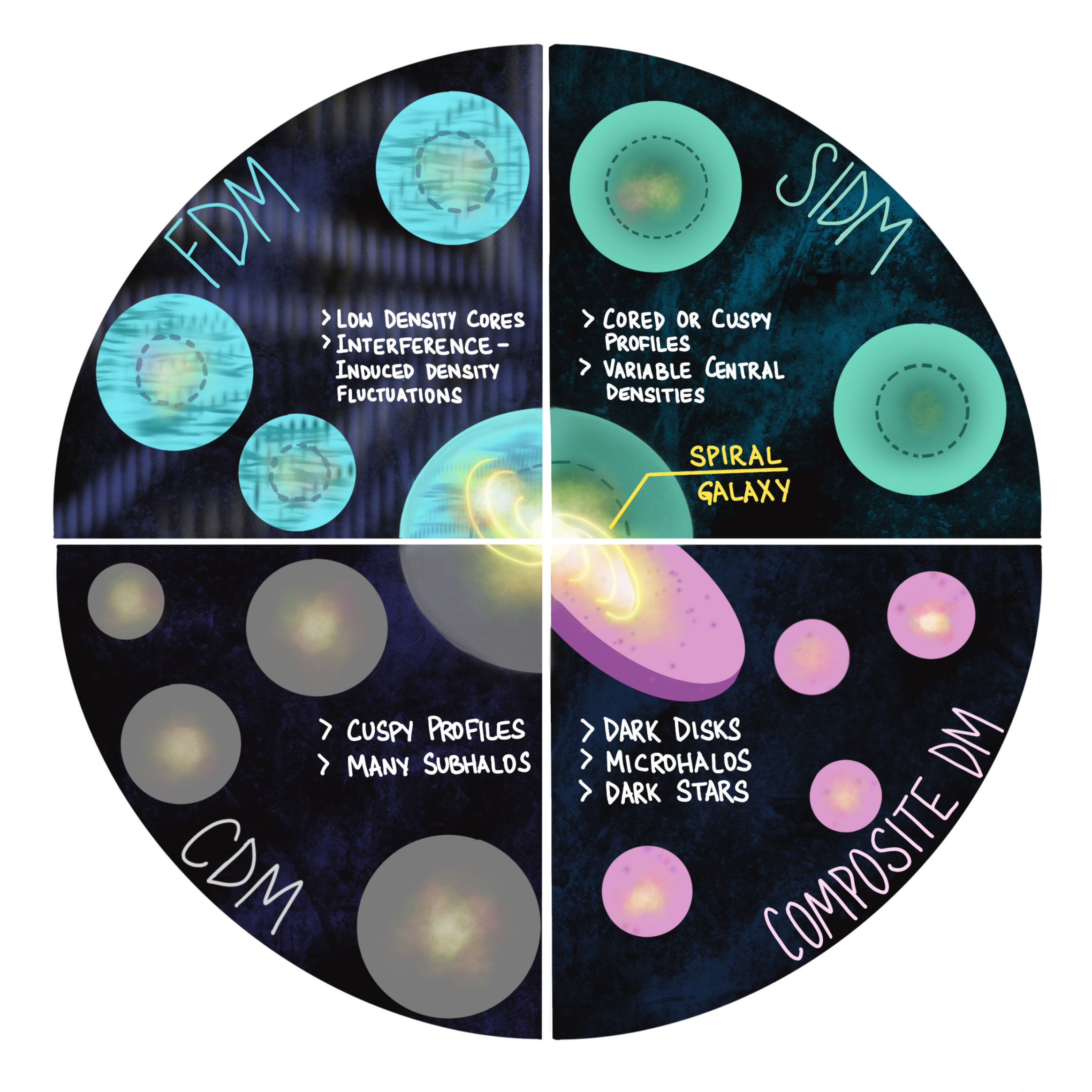}
    \caption{A schematic depiction of different \emph{in situ} DM microphysics effects on the morphology of DM halos and subhalos. See Sections~\ref{insitu} and \ref{sec:hybrid_mods} for further discussion.}
    \label{fig:insitu}
\end{figure}
\subsubsection{Self-interacting dark matter}
\label{sec:SIDM}

\begin{figure}[h]
\centering
\includegraphics[scale=0.4]{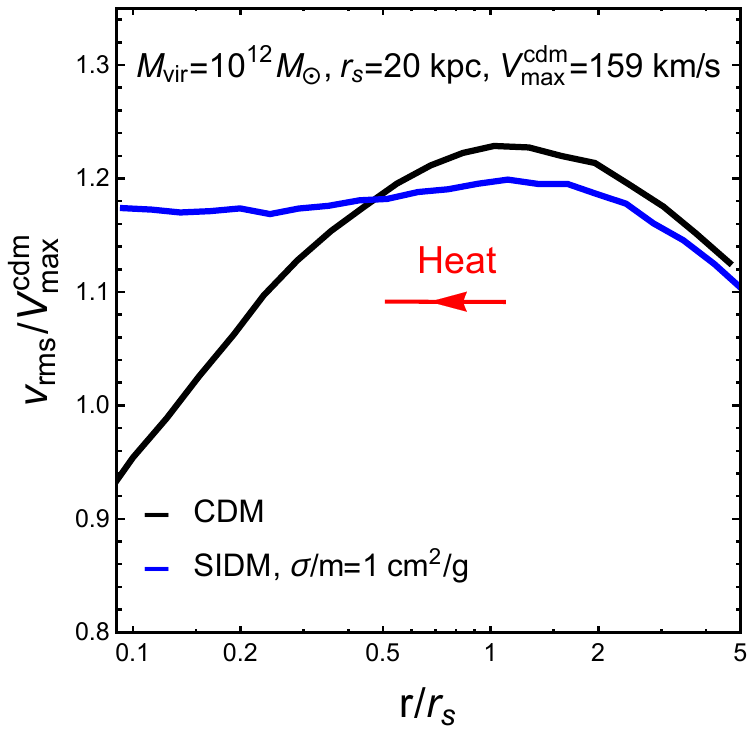}
\includegraphics[scale=0.4]{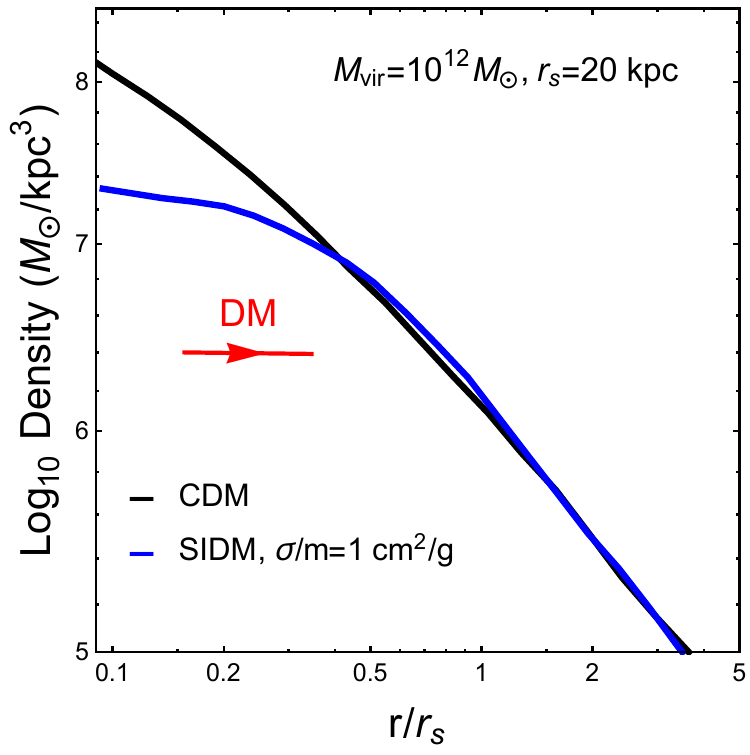}
\includegraphics[scale=0.4]{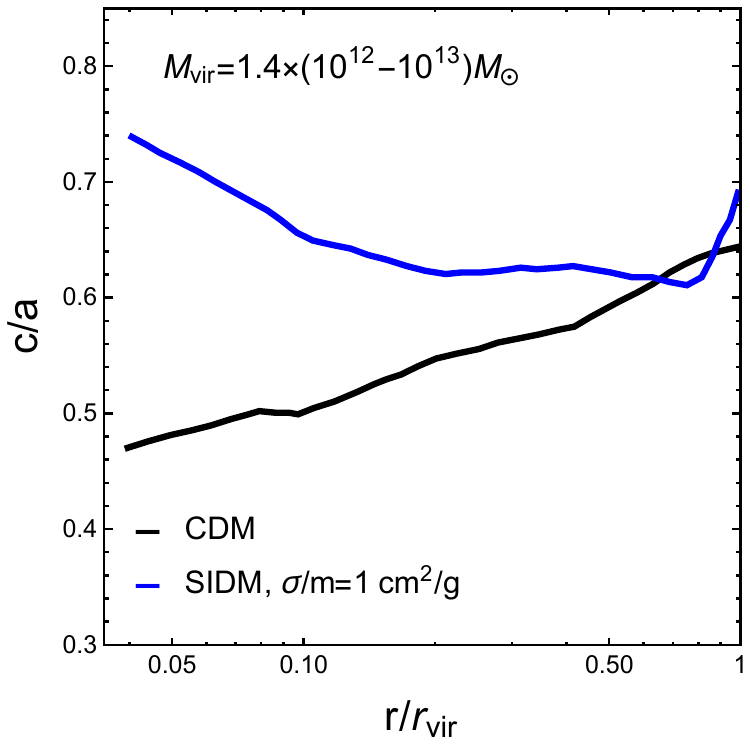}
\caption{Profiles of dark matter velocity dispersions (left), densities (middle), and shapes (right) predicted in CDM (black) and SIDM (blue). Taken from~\cite{Tulin:2017ara} with the simulation data in~\cite{Rocha:2012jg}.\label{fig:sidm}
}
\end{figure}

In the prevailing CDM theory, DM is assumed to be cold and collisionless. CDM is extremely successful for the large-scale structure of the Universe, corresponding to distances greater than $\mathcal{O}({\rm Mpc})$ today. On smaller scales, cosmological simulations have provided several predictions for the structure of CDM halos and abundance of substructures on these scales. However, it remains unclear whether these predictions are all consistent with observations. In particular, DM distributions in galactic halos are much more diverse than predicted in CDM; see~\cite{Bullock:2017xww} for a review on the small-scale issues of CDM. 
 
A promising alternative to CDM is self-interacting DM (SIDM), see~\cite{Tulin:2017ara} for a review. In this scenario, DM particles scatter with each other through $2 \to 2$ interactions. Collisions conduct heat from the hotter outer to the cooler inner regions of a DM halo, leading to a radical deviation from CDM predictions for the inner halo structure~\cite{Spergel:1999mh,Vogelsberger:2012ku,Rocha:2012jg,Peter:2012jh}. Fig.~\ref{fig:sidm} shows profiles of SIDM velocity dispersions (left), densities (middle) and shapes (right), with (blue) and without (black) DM self-interactions. Compared to CDM, the self-interactions thermalize the inner halo, and produce a shallower density profile and a rounder shape, which can be probed in astrophysical observations over a wide range of galactic mass scales, see Sec.~\ref{sec:cluster} for more details.

Recent studies show that SIDM predicts more diverse DM distributions than CDM in galactic systems. For spiral galaxies in the field, SIDM thermalization ties DM and baryonic distributions together and the inner halo density could be either cored or cuspy, depending on the baryon concentration~\cite{Kaplinghat:2013xca}. This feature helps explain diverse galactic rotation curves of the spirals~\cite{Kamada:2016euw,Ren:2018jpt}, a long-standing challenge for CDM, see Sec.~\ref{sec:rotation}. For the substructures, the interplay between self-interactions and tidal interactions can change halo properties in a dynamical way, resulting in a large spread in the central density of subhalos~\cite{Nishikawa:2019lsc,Sameie:2019zfo,Kahlhoefer:2019oyt,Turner:2020vlf,Yang:2021kdf,Zeng:2021ldo}. For example, SIDM may naturally explain diffuse stellar distributions and low DM contents of newly-discovered ultra-diffuse galaxies~\cite{Carleton:2019,Yang:2020iya}. 

In addition, SIDM predicts gravothermal catastrophe~\cite{Balberg:2002ue}, i.e., a phase of gravothermal evolution of the halo at which the central density becomes much higher than its CDM counterpart. For a typical halo, the timescale for the onset of the collapse is longer than the Hubble time. However, the process can be accelerated if the self-interactions have a dissipative component~\cite{Essig:2018pzq,Huo:2019yhk}, the halo is in the tidal environment~\cite{Nishikawa:2019lsc,Sameie:2019zfo}, or the baryon concentration is high~\cite{Elbert:2016dbb,Sameie:2018chj}, leading to unique signatures in astrophysical observations. Refs.~\cite{Balberg:2001qg,Pollack:2014rja,Choquette:2018lvq,Feng:2020kxv,Xiao:2021ftk,Feng:2021rst} further show that SIDM could explain the origin of supermassive black holes via the mechanism of gravothermal collapse.

The figure of merit for the self-interactions is the cross section per unit mass, $\sigma/m$. To have an observable effect on DM halos over cosmological timescales, the required $\sigma/m$ must be at least
\begin{equation}
\sigma/m \sim 1 \; {\rm cm^2/g}\approx 2~{\rm barns/{\rm GeV}}\approx 2 \times 10^{-24} \; {\rm cm^2/GeV}
\end{equation}
in dwarf galaxies, where characteristic velocities of DM particles are $v\lesssim100~{\rm km/s}$. On the other hand, observations of galaxy clusters put an upper limit of $\sigma/m\lesssim0.1~{\rm cm^2/g}$ for $v\sim1000~{\rm km/s}$~\cite{Tulin:2017ara,Andrade:2020lqq}. Thus the self-scattering cross section is necessarily velocity-dependent. From a particle physics perspective, the relevant SIDM cross section is many orders of magnitude larger than the typical weak-scale cross section expected for a WIMP. Evidence for self-interactions would therefore point toward a new dark mediator particle that is much lighter than the weak scale~\cite{Feng:2009mn,Feng:2009hw,Buckley:2009in,Loeb:2010gj,Tulin:2013teo,vandenAarssen:2012vpm,Kaplinghat:2015aga,Cyr-Racine:2015ihg}. The mediator must decay to massless degrees of freedom to avoid over closing the Universe. If it couples to the SM particles, SIDM could produce signals in terrestrial DM detection experiments~\cite{Kaplinghat:2013yxa}. On the other hand, if the mediator couples to massless particles in the dark sector, SIDM predicts dark acoustic oscillations, leading to a suppressed matter power spectrum~\cite{Boehm:2000gq,vandenAarssen:2012vpm,Cyr-Racine:2012tfp,Cyr-Racine:2015ihg,Huo:2017vef}.

\subsection{Hybrid Impact on Structure Formation}\label{sec:hybrid_mods}


\subsubsection{Composite dark matter}
\label{subsec:composite}

For hints that the dark matter may not be described by the simplest model, one need look no further than the Standard Model (SM) of particle physics, which is very far from minimal. The multitude of SM particles and forces give rise to a large variety of microscopic phenomena, such as composite hadronic and nuclear bound states as well as dissipation, atoms and chemistry. We probably would never have expected the complexity of the SM if we did not live in it. What other strange sectors could be making up a substantial fraction of the Universe's matter budget? This bottom-up plausibility argument sets the stage for considering \emph{dark complexity}, the possibility of hidden sectors with particles and forces that approach the SM in complexity, with possibly equally interesting dynamics in the early Universe and today. 

Dark complexity and the theories that can be included in this umbrella term, such as \emph{composite dark matter}~\cite{Cline:2021itd,SpierMoreiraAlves:2010err,
Boddy:2014yra,
Kribs:2016cew,
Antipin:2015xia,
Cline:2013zca,
Khlopov:2008ty,
Kribs:2009fy,
Alves:2009nf,
Ko:2017uyb,Tsai:2020vpi} and \emph{atomic dark matter}~\cite{Goldberg:1986nk,Kaplan:2009de, Kaplan:2011yj,Cline:2012is,Cline:2013pca,Fan:2013yva,Fan:2013tia,Fan:2013bea,Cyr-Racine:2013fsa,Boddy:2016bbu,Rosenberg:2017qia,Ghalsasi:2017jna,Gresham:2018anj,Essig:2018pzq,Alvarez:2019nwt,Cline:2021itd,Cyr-Racine:2021alc,Blinov:2021mdk}, are also deeply motivated from a top-down point of view. 
Many BSM theories postulate the existence of hidden sectors related by discrete symmetries to the SM~\cite{Chacko:2005pe,Barbieri:2005ri,Chacko:2005vw,Chacko:2016hvu,Craig:2016lyx,Chacko:2018vss,Chacko:2021vin,GarciaGarcia:2015pnn,Foot:2002iy,Foot:2003jt,Berezhiani:2003xm,Foot:1999hm,Foot:2000vy,Foot:2003eq, Foot:2004pa}. 
This includes models of ``neutral naturalness'' such as the Mirror Twin Higgs that address the hierarchy problem of the Higgs boson mass \cite{Chacko:2005pe,Barbieri:2005ri,Chacko:2005vw,Chacko:2016hvu,Craig:2016lyx,Chacko:2018vss,Chacko:2021vin,GarciaGarcia:2015pnn,Harigaya:2019shz} by positing the existence of a hidden sector related to the SM by a $\mathbb{Z}_2$ or other discrete symmetry, resulting in analogues of electromagnetism, strong and weak forces as well as various matter states existing in the dark sector. Interestingly, while these dark sectors are qualitatively analogous to the SM in their ingredients, their dynamics can be very different owing to different effective values of the corresponding fundamental constants.
This can result in a fraction of dark matter being made up of dark baryons, with dark protons, dark electrons, dark photons, and even dark nuclear forces that are similar to their SM counterparts but with possibly different values for masses and interaction strengths. Another solution to the hierarchy problem that may give rise to similar phenomena is $N$-Naturalness~\cite{Arkani-Hamed:2016rle}. 

Composite DM~\cite{Cline:2021itd,SpierMoreiraAlves:2010err,
Boddy:2014yra,
Kribs:2016cew,
Antipin:2015xia,
Cline:2013zca,
Khlopov:2008ty,
Kribs:2009fy,
Alves:2009nf,
Ko:2017uyb,Tsai:2020vpi} is a well-motivated DM scenario whereby the hidden sector is some analogue of QCD that could be as simple as  some dark fermion(s) charged under a confining dark gauge force. All or part of DM could then be made up of a dark-hadronic bound state, with the details of its possible dynamics in our Universe today ranging from being effectively collisionless to acting like self-interacting dark matter~\cite{Cline:2013zca,Boddy:2014yra,Ko:2017uyb,Ibe:2018juk,Tsai:2020vpi} (see Section~\ref{sec:SIDM}) or even the nucleonic part of atomic dark matter. As a result, composite DM can realize many rich signatures across the broad range of halo masses discussed in this white paper.

An archetypical scenario that leads to \emph{dissipation} in DM halos is \emph{atomic dark matter}~\cite{Goldberg:1986nk,Kaplan:2009de, Kaplan:2011yj,Cline:2012is,Cline:2013pca,Fan:2013yva,Fan:2013tia,Fan:2013bea,Cyr-Racine:2013fsa,Boddy:2016bbu,Rosenberg:2017qia,Ghalsasi:2017jna,Gresham:2018anj,Essig:2018pzq,Alvarez:2019nwt,Cline:2021itd,Cyr-Racine:2021alc,Blinov:2021mdk}, which 
postulates that dark matter contains at least two states with different masses and opposite charge under a \emph{dark electromagnetism} $U(1)_D$ gauge symmetry. Straightforward generalizations include different charge ratios of the ``dark nucleon'' and the ``dark electron,'' and the possibility of many dark nucleon species. The latter scenario is especially motivated if atomic dark matter is realized together with composite DM, i.e. if the dark nucleons are actually dark-hadronic bound states which in turn form different nuclei, in analogy to the SM. The dark photon could be detected in future CMB-S4 and CMB-HD measurements~\cite{CMB-S4:2016ple,Sehgal:2019ewc}, and the long-range dark-electromagnetic interaction also causes dark acoustic oscillations of the atomic DM component (see e.g.~\cite{Cyr-Racine:2013fsa, Chacko:2018vss}), which can leave oscillatory deviations in Large Scale Structure matter power spectrum measurements compared to the $\Lambda$CDM expectation. However, by far the most drastic signatures of atomic dark matter and dissipation occur at galactic and smaller scales. They could easily be compatible with cosmological and self-interaction bounds, especially if the dark atoms make up a $\lesssim \mathcal{O}(10\%$) subcomponent of dark matter~\cite{Fan:2013yva,Cyr-Racine:2013fsa,Chacko:2018vss}. Like their SM counterparts, dark particles could cool and form structure, leading to the formation a dark disk or dark microhalos~\cite{Fan:2013yva,Fan:2013tia,Ghalsasi:2017jna, Ryan:2021tgw, Gurian:2021qhk, Ryan:2021dis}, which leave an observable gravitational imprint on visible matter that can be observed for instance using astrometry~\cite{Kramer:2016dqu,Schutz:2017tfp,Buch:2018qdr,Widmark:2021gqx}, as discussed further in Section~\ref{stellardynamics}.

On even smaller astrophysical scales, collapsing atomic dark matter would form mirror stars~\cite{Foot:1999hm, Foot:2000vy,Curtin:2019lhm, Curtin:2019ngc}, possibly emitting dark photons long past their Kelvin--Helmholz time if dark nuclear interactions release energy in their cores. This opens a new window for astrophysical dark matter searches, since it has recently been shown~\cite{Curtin:2019lhm, Curtin:2019ngc} that mirror stars can produce electromagnetic signals that may be visible in optical/IR and X-ray telescopes if the dark photon has a tiny kinetic mixing with the SM photon. Such a small mixing $\epsilon F_{\mu \nu} F_D^{\mu \nu}$ is a renormalizable operator in the Lagrangian and generically expected to be produced at some loop level in the complete theory~\cite{Gherghetta:2019coi}, but even without such a coupling, the gravitational waves of mirror neutron star mergers will be observable~\cite{Hippert:2021fch}.
In the absence of dark nuclear interactions, atomic dark matter can collapse directly to black holes~\cite{Ryan:2021tgw, Gurian:2021qhk, Ryan:2021dis}  (with very different masses than expected from SM astrophysics), leading to detectable gravitational waves in their mergers~\cite{Singh:2020wiq, Shandera:2018xkn}.
The Vera Rubin telescope will also be able to detect the microlensing signal of mirror stars in our galaxy~\cite{Winch:2020cju}. 
Finally, detection of dissipative atomic dark matter in our stellar neighborhood either via ground-based detectors~\cite{Chacko:2021vin} or anomalous cooling of white dwarfs~\cite{Curtin:2020tkm} depends on its distribution in the galaxy and the degree to which it has undergone cooling, recombination, and collapse into a dark disk. 
A positive detection in any combination of the abovementioned channels would therefore not only reveal the existence of dark matter, but allow us to directly probe details of its distribution within our galaxy, which would reveal aspects of hidden sector dynamics that cannot be probed in other experiments or observations.

\subsubsection{Low-mass bosons}\label{sec:FDM}

One hypothesis postulates that dark matter halos are comprised
of ultralight bosons ~\cite{1983PhLB..122..221B,1994PhRvD..50.3650S,2000PhRvL..85.1158H}, which are broadly expected to arise in the context of many string theories~~\cite{witten1984some,svrcek2006axions,Arvanitaki:2009fg,acharya2010m,Cicoli:2012sz}. This so-called Fuzzy Dark Matter (FDM) model
introduces wave dynamics in the dark matter
on the scale of the de Broglie wavelength due to the extraordinarily high particle occupation number. 
Lately, this idea has seen increased attention because bosons with a mass of $\sim10^{-22}~{\rm eV}$ have a de Broglie wavelength on kiloparsec scales and the resulting ``quantum pressure'' (which arises due to the macroscopic de Broglie wavelength and not because of additional forces) affects the galaxy-scale distribution of dark matter, which may have an impact on the
``small-scale'' challenges of the CDM paradigm \cite{2015PNAS..11212249W,2017ARA&A..55..343B}.
Numerical simulations testing the nonlinear structure formation and evolution of this dark matter model have recently become possible \cite{2014NatPh..10..496S,2017MNRAS.471.4559M,2019PhRvL.123n1301M,2020MNRAS.494.2027M,2018PhRvD..97f3507D,2020PhRvD.101h3518V,2020PhRvD.102h3518S,2020PrPNP.11303787N,2021MNRAS.501.1539N,2021PhRvD.103b3508L}, 
placing constraints on the particle mass and making predictions for unique small-scale features of the model.
The arising quantum pressure suppresses small-scale power in the initial dark matter power spectrum \cite{2000PhRvL..85.1158H, Hlozek:2014lca},
modifies the halo mass function 
\cite{Marsh:2018zyw,2020PhRvD.101l3026S}, creates extended low-density soliton cores at the centers of dark matter halos
\cite{2014NatPh..10..496S,2015MNRAS.451.2479M}, exhibits transient density fluctuations on very small scales as a result of wave interference which can dynamically heat stellar streams and clusters~\cite{Marsh:2018zyw,Amorisco:2018dcn,Church:2018sro,Dalal:2020mjw}, and induces nontrivial dynamical friction on subhalos~\cite{Lancaster:2019mde}. A challenge for the FDM model continues to be whether an ultralight particle mass can simultaneously be consistent with the Lyman-$\alpha$ forest power spectrum extracted from high-redshift quasars~\cite{2017PhRvL.119c1302I,2019MNRAS.482.3227N} as well as the subhalo mass function~\cite{2020PhRvD.101l3026S, Nadler:2020prv}
and still simultaneously explain the core sizes of satellite galaxies
\cite{2020ApJ...893...21S,2020ApJ...904..161B}. The inclusion of sufficient baryonic physics in a realistic cosmological setting~\cite{Mocz:2019pyf,Mocz:2019uyd,Veltmaat:2019hou} will also be necessary in determining the observable effects of FDM. In some cases, these effects can enhance rather than suppress structure beyond the FDM Jeans scale.

There are still many outstanding theoretical challenges concerning the origin and non-gravitational interactions of FDM. A candidate particle for the ultralight boson is the axion, which would arise from a symmetry-breaking needed to solve the strong CP problem \cite{1977PhRvL..38.1440P,1978PhRvL..40..223W} (see the discussion in Section~\ref{sec:non-therm}). This specific model would also give rise to a weak attractive self-interaction which may further alter the dark matter distribution. Non-gravitational self-interactions can lead to a range of phenomenological consequences including the formation and dynamics of solitons, oscillons, oscillatons, axitons, axion-stars, and vortices in the early and present-day Universe \cite{2018PhRvD..97b3529D,Kolb:1993hw,Schive:2014dra,Levkov:2018kau,Amin:2019ums,Hui:2020hbq,Arvanitaki:2019rax,Desjacques:2017fmf,Leong:2018opi,Kirkpatrick:2021wwz,Glennon:2020dxs}. Additionally, due to terms in the potential that are not quadratic in the axion field,
the axion energy density does not necessarily redshift entirely as DM. Current bounds from the Cosmic Microwave Background set an approximate limit on the axion decay constant, $f_a \gtrsim 10^{13} {\rm GeV} (10^{-20}~{\rm eV}/m_a)$ \cite{Dror:2020zru}, independently of the axion production mechanism. All of these effects are just beginning to be explored, and many further theoretical developments can be anticipated in the coming decade.

\section{Cluster Scale}
\label{sec:cluster}

\subsection{Scale and Objects}

Clusters of galaxies are the largest ($M \gtrsim 10^{13.5} M_\odot/h$) gravitationally bound objects in the Universe, with DM forming most of their mass. Detailed observations of galaxy clusters provided some of the earliest evidence for the existence of DM (see e.g.~Ref.~\cite{Zwicky:1933gu}), and these objects remain excellent laboratories to probe the collisional properties of DM particles. Perhaps the best known example of such an object is the \emph{Bullet Cluster} \cite{2004ApJ...606..819M,Randall:2007ph,2017MNRAS.465..569R}, for which a clear separation between the hot gas (probed via X-ray emission) and the bulk of the mass of the two merging sub-clusters (probed via gravitational lensing) can be used to constrain the interaction cross section between DM particles at high velocities. Beyond merging clusters, detailed measurements of the density profile of relaxed clusters and of the position of their brightest central galaxy away from the cluster's center could also detect the presence of central DM cores created by DM self-interaction. 

\subsection{Observational Probes}
In the following, we outline several observational aspects of galaxy clusters that display sensitivity to DM physics.

\subsubsection{Density profiles of clusters}\label{subsec:dens_clus}

Through a combination of data from optical, CMB, and X-ray surveys, it is now possible to accurately measure the stacked weak lensing profiles around massive galaxy clusters ($M \gtrsim 10^{13.5} M_\odot/h$)~\cite{2018ApJ...864...83C,2021MNRAS.507.5758S}.
\begin{wrapfigure}{R}{0.55\textwidth}
\includegraphics[width=0.42\textwidth]{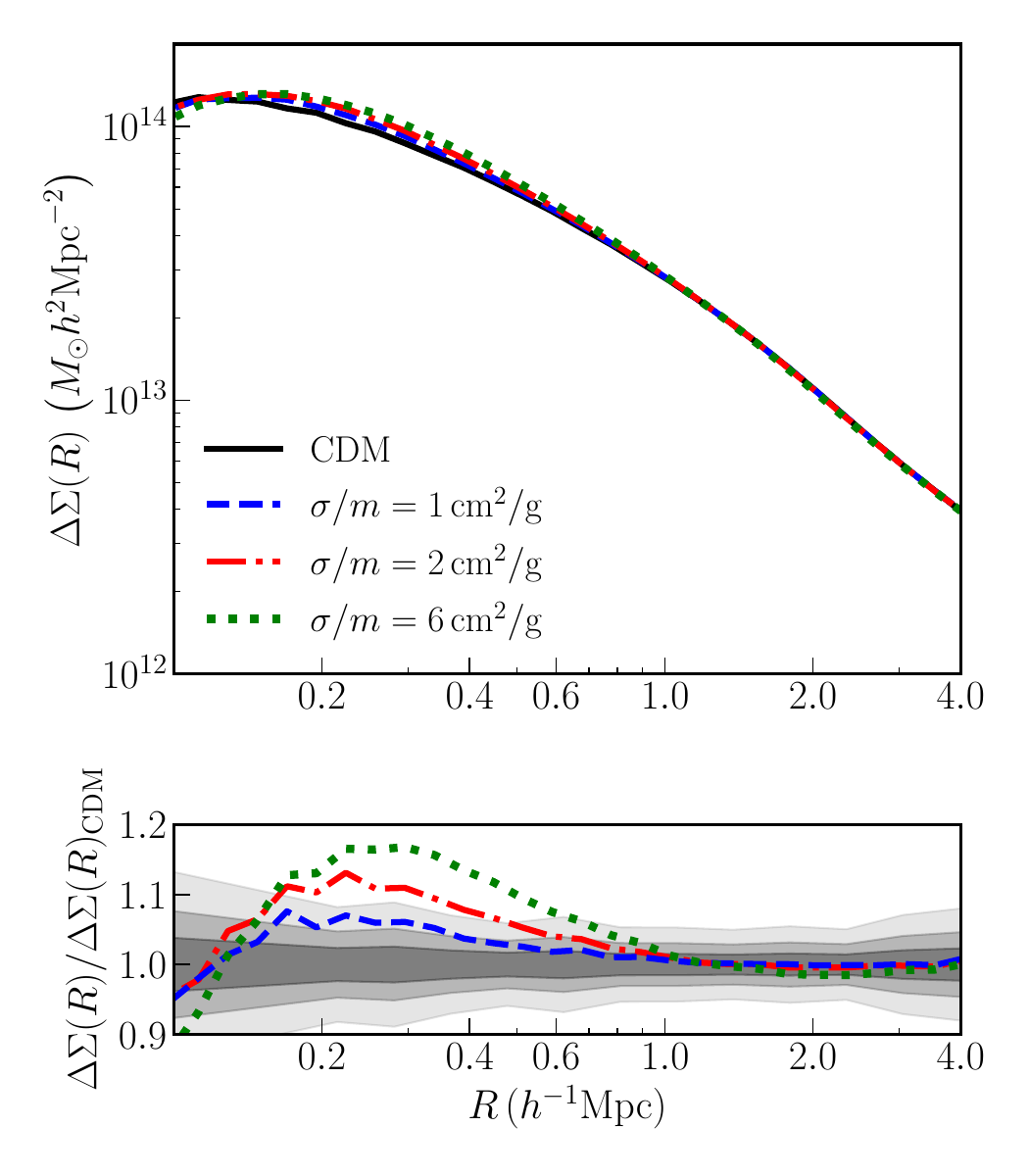}
\centering
\caption{Effect of SIDM on the lensing signal from clusters in the mass range $1-2 \times 10^{14}M_\odot/h$  (top panel). The residuals from the CDM prediction is plotted in the bottom panel. The lightest grey shaded region in the bottom panel corresponds to the measurements error on the lensing profile of a similar sample from the DES Y1 data. The progressively darker shaded regions correspond to predictions for the measurements errors expected in DES Y3 and LSST at Rubin Observatory. These error forecasts only take into account the increased sky coverage, and not expected improvements in the depth and photometry.}
\label{fig:cluster_lesning_forecast}
\end{wrapfigure}
The signal-to-noise is expected to increase significantly in the next decade as the next generation optical (e.g.~the Legacy Survey of Space and Time (LSST) at the Rubin Observatory \cite{LSST:2008ijt}) and microwave (e.g.~CMB-S4 \cite{CMB-S4:2016ple}) surveys come online. These will allow for measurement errors at the percent level from scales comparable to the scale radius of the clusters out to the cluster outskirts. The lensing signal is sourced by the density profile of all matter, including the dark matter component, of the clusters. Therefore, the precise weak lensing measurements around massive clusters can be used to constrain those models of dark matter physics that can affect the density profiles, especially on scales of $\sim 0.2 h^{-1}{\rm Mpc}$ to $\sim 2.0 h^{-1}{\rm Mpc}$.

Self-interacting dark matter (SIDM) models (see Section \ref{sec:SIDM}) are known to produce characteristic signatures on the density profiles of halos. For interaction cross sections per DM mass of $\sigma/m \sim 1 \,{\rm cm}^2/{\rm g}$ at the characteristic velocity scale set by the halo mass of interest, the most prominent signature of SIDM is a central core. While the size of the core itself is typically smaller than the scale radius, the radial profile is affected all the way out to the virial radius due to rearrangement of particles driven out of the core by the interactions. For clusters, therefore, the weak lensing measurements on scales of $\sim 0.2 h^{-1}{\rm Mpc}$ to $\sim 2.0 h^{-1}{\rm Mpc}$, are a sensitive probe to SIDM cross sections at velocity $v\sim 1000\, {\rm km/s}$~\cite{2020JCAP...02..024B}. Existing data from surveys, e.g. Year 1 data from the Dark Energy Survey (DES) has sufficient statistical power to rule out cross sections above $\sim 2\, {\rm cm}^2/{\rm g}$, similar to that obtained from the study of the Bullet Cluster (see, e.g.,~Refs.~\cite{2004ApJ...606..819M, 2017MNRAS.465..569R}). This is shown in Fig.~\ref{fig:cluster_lesning_forecast}, which is adapted from Ref.~\cite{2020JCAP...02..024B}. The top panel shows the effects of SIDM on the lensing profile of clusters in the mass range $1 - 2\times 10^{14}M_\odot/h$. The bottom panel shows the residuals of the SIDM models compared to the CDM model. The light grey shaded region in the bottom panel represents the measurement error on a similar cluster sample from DES Y1 data~\cite{2018ApJ...864...83C}. The progressively darker bands represent the expected measurement errors from DES Y3 data and LSST at VRO, obtained by rescaling the DES Y1 error bars by an appropriate factor to take into account the increased sky area. These projections are, therefore, conservative in the sense that they do not take into account the greater depth, and therefore more background sources, compared to DES Y1. With LSST, the statistical power is sufficient to constrain the cross section at better than $\sigma/m = 0.5\,{\rm cm}^2/{\rm g}$ at these large velocities.

Since these measurements are determined by the cluster outskirts, they are less affected by baryonic effects than measurements closer to the cluster center. However, in light of the expected statistical constraining power on SIDM models expected in the future, any possible effects of baryonic effects on the size or shape of the signal need to be modeled to reduce the systematic errors associated with the interpretation of these measurements in terms of SIDM cross sections.

\subsubsection{Merging clusters}

If DM particles scatter with one another at astrophysically important rates (i.e.~if DM is some form of SIDM, see Section~\ref{sec:SIDM}), then two likely scenarios are either that the cross-section is independent of velocity, or that it decreases with increasing velocity. In either of these scenarios merging galaxy clusters are interesting, because of the high relative velocities between DM particles. In the case of a velocity-independent cross-section, the rate of scattering per particle is proportional to the product of the density and the mean pairwise velocity between dark matter particles. The high densities and velocities in merging clusters then mean that these systems would have high rates of SIDM scattering, making them a useful laboratory for these interactions. On the other hand, if DM self-interactions have a velocity-dependent cross-section, merging clusters would then provide complementary information to what can be learnt from lower-mass systems \cite{2016PhRvL.116d1302K}, because they measure the cross-section at an otherwise inaccessible velocity scale. 

In addition to the interesting velocity scale at which merging clusters probe the SIDM cross section ($v\sim$1000--3000 km$\,$s$^{-1}$), another unique feature of merging clusters is that they are systems with a clear directionality. This means that the angular dependence of the SIDM scattering cross-section can be probed \cite{2014MNRAS.437.2865K, 2017MNRAS.467.4719R, 2021MNRAS.505..851F}. This is not the case for more isotropic effects of SIDM, such as core formation at the centre of halos, where SIDM models with different forms of differential cross section can lead to the same evolution of the density profile \cite{2017MNRAS.467.4719R}.

The archetypal cluster merger is the Bullet Cluster, in which a low-mass ($\sim 10^{14} \msun$) cluster (called ``the bullet'') has passed through a more massive ($\sim 10^{15} \msun$) cluster, with the collision taking place approximately in the plane of the sky \cite{2002ApJ...567L..27M}. This system contains a textbook example of a bow shock (visible in X-rays), from which the speed and direction of the collision can be inferred. In addition, strong and weak gravitational lensing can be used to map the distribution of dark matter within the merging cluster system \cite{2004ApJ...604..596C, 2006ApJ...652..937B}, from which it can be inferred that the DM must be at least approximately collisionless in order for it to be leading in front of the (highly collisional) hot gas, which is slowed down by hydrodynamical forces. This statement that DM must be approximately collisionless can be turned into a more formal constraint on the SIDM cross section by considering the optical depth for SIDM interactions as particles from the bullet halo pass through the main halo \cite{2004ApJ...606..819M}. The projected DM density in the inner regions of the main halo is approximately $\Sigma = 0.2 \, \mathrm{g \, cm^{-2}}$, and the main halo would become optically thick for SIDM interactions if $\tau = \Sigma \, \sigma/m > 1$. Therefore, requiring that the main halo is not optically thick for SIDM interactions implies $\sigma/m < 5 \, \mathrm{cm^{2} \, g^{-1}}$. In other words, for cross sections $5 \, \mathrm{cm^{2} \, g^{-1}}$ or larger, most bullet halo particles would have scattered with a main halo particle during the collision, and the DM's behaviour would not be approximately collisionless.

The SIDM constraint from the Bullet Cluster was tightened somewhat by comparison with $N$-body simulations of merging clusters with different SIDM cross sections \cite{2008ApJ...679.1173R}, but further attempts to refine this have struggled to lower the upper-limit on the SIDM cross section \cite{2014MNRAS.437.2865K, 2017MNRAS.465..569R, 2017MNRAS.469.1414K}. Even finding one Bullet Cluster-like system in the observed Universe was fairly unlikely \cite[e.g.][]{2015JCAP...04..050K} so it is highly likely that any merging systems yet to be discovered will be less extreme in terms of their mass and collision velocity, and therefore in terms of the expected size of SIDM effects.

One potential avenue for improving constraints is to study a large number of more minor mergers \cite{2011MNRAS.413.1709M}, but an initial attempt at doing this \cite{2015Sci...347.1462H} was called into question when individual systems from this sample were analysed in more detail \cite{2018ApJ...869..104W}, and it was found that the offsets between the different components (DM, gas and stars) changed considerably when using more comprehensive data on the same clusters. LSST, Euclid, and Roman will improve the quality of weak and strong lensing constraints across much of the sky, and telescopes such as the Super-pressure Balloon-borne Imaging Telescope (SuperBIT) can provide targeted observations that will allow individual systems to be explored in more detail. Nevertheless, simply improving the quality of lensing measurements of merging systems will not significantly improve constraints on SIDM without some theoretical advances in how to extract DM physics from observed mergers.

\subsubsection{Brightest cluster galaxy oscillations}

As mentioned in Section \ref{subsec:dens_clus}, SIDM generally predicts that dark matter halos should have constant density cores. In galaxy clusters, the largest bound gravitational objects in the Universe, these cores are expected to be large, $\sim$100 kpc, for a velocity-independent cross section of $\sigma/m \sim$ 1 cm$^2$/g \citep{rocha2013}.  As dynamical friction does not operate in the flat potential of cored profiles, one can imagine that if two clusters of relatively similar mass merge, their galaxies will ``slosh,'' or oscillate, in the core of the merger remnant.  In particular, the brightest cluster galaxy (BCG) should oscillate within the merger remnant on an orbit determined by the size of the core.  The degree to which BCGs are offset from the centers of relaxed clusters can thus constrain the self-interaction cross section.  As the core size is a strong function of cross section, this is a potentially sensitive test of $\sigma/m$.  BCGs are observed to be offset from their parent clusters' X-ray center by $\sim$10 kpc \citep{lauer2014}, which implies a cross section of order $\sim$0.1 cm$^2$/g \citep{kim2017}.  A follow-up study of 10 relaxed clusters observed with the Hubble Space Telescope similarly indicated an oscillation amplitude of 11.82$^{+7.3}_{-3.0}$ kpc from their mass peaks inferred from strong lensing \citep{harvey2017}.

Several theoretical questions remain.  Initial constraints were based on dark-matter-only simulations of idealized equal mass mergers.  The inclusion of gas has been shown to produce smaller SIDM cores and reestablish a potential gradient \citep{robertson2018}.  This can reintroduce dynamical friction and allow the orbits of the BCGs to decay.  Simulations exploring non-equal mass mergers are also key.  Equal mass mergers are expected to be much rarer than unequal mass mergers, as hinted by the observed BCG offset probability distribution, which is not consistent with that expected for a population dominated by equal mass mergers.  In unequal mass mergers, the BCG of the more massive cluster (and presumably the BCG of the merger remnant) would not be significantly perturbed, resulting in smaller-scale oscillations.  Simulations also have yet to explore the effect of velocity-dependent cross sections.  As such models tend to favor low cross sections at cluster-scale velocities, oscillations are likely small in such models as well.

Observationally, new weak lensing mass maps from upcoming surveys with instruments such as the {\it Roman} and {\it Euclid} space telescopes, as well as SuperBIT will allow constraints to be improved by a factor of at least two \citep{harvey2019}. Positional offsets can be complemented with measurements of oscillations in velocity space, which are also expected to be large, and observable with the Maunakea Spectroscopic Explorer \citep{The_MSE_Science_Team2019}.

\subsubsection{Substructure in massive clusters}

As in the case of the Milky Way, the spatial and velocity distribution of substructure in galaxy clusters is sensitive to DM physics. In the presence of SIDM with a velocity-dependent interaction cross section, the cluster subhalos and the main cluster halo probe different regions of the velocity space. Indeed, while scattering between subhalo DM particles occurs at typical galactic speed ($v\sim100-300$ km/s), collisions between main host and subhalo DM particles occur at the typical velocities found in clusters, which tend to be significantly higher. In these systems, the interaction cross section at galactic velocities will be responsible for core formation or even the onset of gravothermal collapse \cite{Nishikawa:2019lsc}, while the DM cross section at the higher cluster velocities would impact the ram-pressure stripping (evaporation) that subhalos experience as they move  within the cluster halo. Thus, detailed observations of cluster substructure using both weak and strong gravitational lensing \citep{Bhattacharyya:2021vyd} could probe the DM cross section over a broad range of velocities \citep{Zeng:2021ldo}. Already, Ref.~\cite{Meneghetti:2020yif} has found a possible discrepancy (see however Ref.~\cite{Bahe:2021bcs}) in the properties of cluster substructure as compared to $\Lambda$CDM predictions, which could potentially be explained in the context of SIDM (see e.g.~Ref.~\cite{Yang:2021kdf}).

\subsection{Theory Connections}
As can be inferred from the above discussion, galaxy clusters are excellent probes of the potential collisional nature of DM in the high-velocity regime. Clusters can be thought as high-energy DM particle colliders, hence providing a unique environment to test for any deviations from the standard CDM scenarios at those energies. Galaxy clusters are thus a key probe of SIDM (see Section \ref{sec:SIDM}), as well as composite DM (see Section \ref{subsec:composite}), which typically have sizable self-interaction. 

\section{Galaxy Scale}

\subsection{Scale and Objects}
Luminous galaxies span a large range of halo masses ($\sim10^8-10^{13}\,M_\odot$), making them very versatile laboratories to probe the physical properties of DM. Observations of galactic rotation curves by Rubin, Ford, and Thonnard \cite{Rubin:1980zd} were instrumental in firmly establishing DM as an integral part of our modern cosmological model. Detailed indirect observations of the DM distribution within galaxies (including within our own) can provide important clues about the fundamental physics of DM. The formation of the first galaxies in the early Universe also holds important information about DM since their abundance is quite sensitive to the low-mass end of the halo mass function.  

\subsection{Observational Probes}
In this section, we outline several probes of dark matter physics on the scales of galaxies. We focus our attention here on ``luminous'' galaxies (including our own), while ultra-faint dwarfs will be treated in Section \ref{sec:beyond_galaxy_threshold}.

\subsubsection{Dwarf galaxy lensing}

\begin{figure}
\centering
\includegraphics[width=14cm]{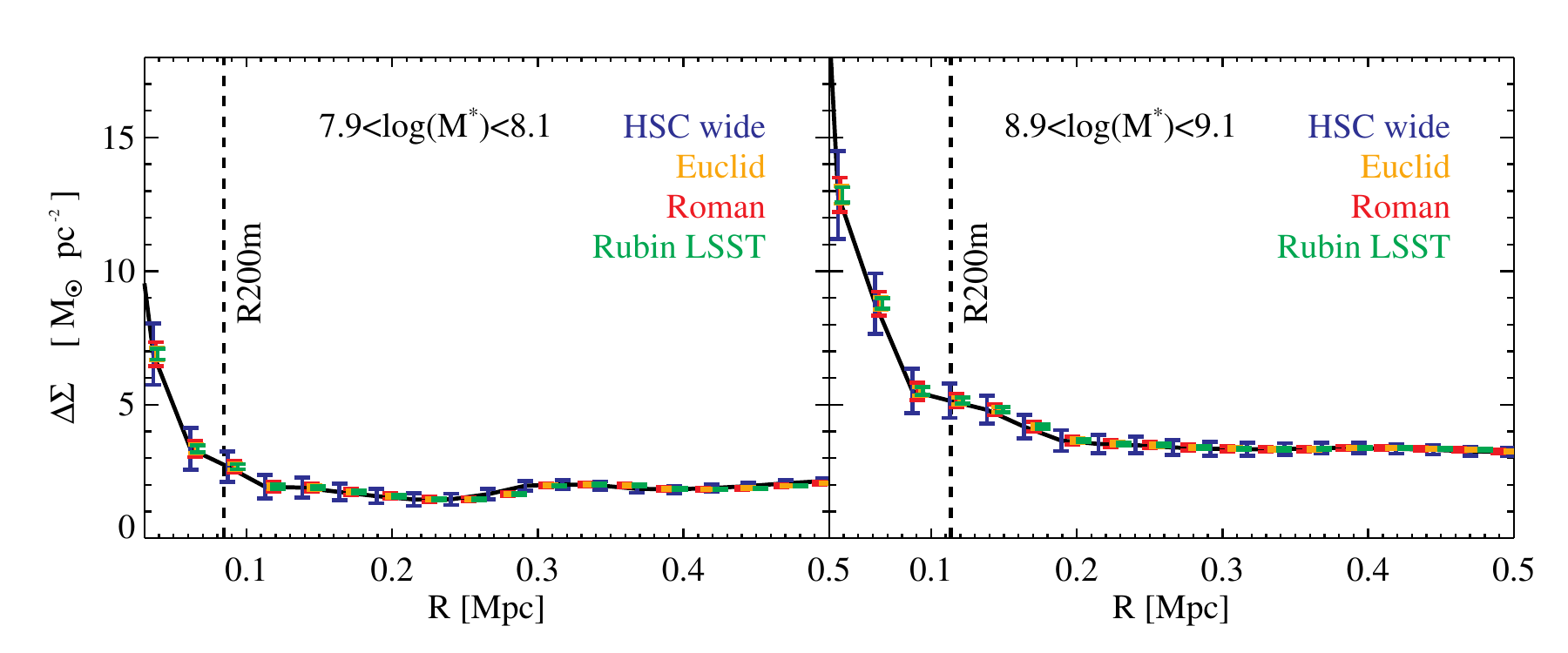}
\caption{Predicted lensing signal and errors for the $\Delta\Sigma$ profile around dwarf lens galaxies. The left panel corresponds to dwarfs with $M_{\star}=10^{8}\,\rm M_{\odot}$ and the right panel corresponds to dwarfs with $M_{\star}=10^{9}\,\rm M_{\odot}$. Errors correspond to a lens samples selected within a narrow mass range (0.2 dex bin width) and for $0<z<0.25$. Predicted diagonal errors are shown for HSC wide survey (blue), \textit{Euclid} (orange), Roman (red), and Rubin LSST (green). Figure taken from Ref.~\cite{Leauthaud2020}.}
\label{lensingdwarfs}
\end{figure}

Dwarf galaxies are a unique probe of the nature of dark matter and the interplay between dark matter and baryonic physics. However, rotation curves of dwarf galaxies only constrain the inner regions of dark matter halos (typically a factor of $\sim$10--20 smaller than the actual halo radius). Any ``halo mass'' estimate from rotation curves is in fact an extrapolation that relies on assumptions about the shape of the dark matter profile (see e.g.~Ref.~\cite{Buckley2018}). One of the most powerful ways to directly probe total halo masses out to the halo radius is ``galaxy--galaxy lensing.'' This is the average weak lensing signal from background ``source'' galaxies around a sample of foreground ``lens'' galaxies. Existing weak lensing measurements have been limited to  $\rm M_{\star}>10^{9}$ $\rm M_{\odot}$.  However, recent work has showed that new generation lensing surveys can measure the lensing signal of $\rm M_{\star}\sim10^{8-9}\,\rm M_{\odot}$ dwarf galaxies with high significance \cite{Leauthaud2020}. Figure~\ref{lensingdwarfs} shows the expected lensing signals for dwarf galaxies from upcoming surveys such as \textit{Euclid}, Roman, and Rubin.
With the number of dwarf lenses available in Rubin LSST, it is even possible to distinguish different density profiles (e.g., cored vs.\ cuspy) using $M_{\star}\sim10^{9-10}\,\rm M_{\odot}$ dwarf galaxies as lenses \cite[Sec.~3.2.1 of Ref.][]{Drlica-Wagner:2019}.
In combination with measurements of galaxy dynamics, lensing measurements for dwarf galaxies will provide unique and novel constraints on non-CDM theories of dark matter and on feedback models at the dwarf scale. To harness the power of large lensing surveys and to tackle this science goal, however, large samples (tens of thousands) of dwarf galaxies with reliable redshift measurements are needed. 

The need for such a large sample precludes directly spectroscopic follow-up \cite{Leauthaud2020}.
Photometric redshifts from \textit{Euclid}, Roman, and Rubin are also unlikely to be accurate enough to reliably select large samples of dwarf galaxies in the mass range $M_{\star}=10^{7-9}\,\rm M_{\odot}$. 
However, there has been much progress in using machine learning algorithms to improve photometric redshifts, especially in the very low-redshift regime \cite{Wu2021,Dey2021}. Combining these novel algorithms with substantial very low-redshift spectroscopic training samples (e.g., SAGA Survey \cite{Mao2021}; DESI LOW-Z Secondary Program, E.~Darragh-Ford et al. in prep.), we can significantly improve sample selection with photometric redshifts in the Rubin LSST era.

In addition to only using photometric redshifts for sample selection, it will soon be possible to select large samples of star-forming dwarfs using narrow-band imaging targeting their strong H$\alpha$ and O$_{\rm III}$ emission lines. As an important pathfinder for this promising approach, the Merian Survey\footnote{\url{https://merian.sites.ucsc.edu}} will use the Blanco telescope to image 867 deg$^2$ of the Hyper-Suprime Camera (HSC) Wide field using two custom-made filters probing H$\alpha$ and O$_{\rm III}$ at $z\sim 0.08$. This survey will provide redshifts to a complete sample of $100,000$ star-forming dwarf galaxies (two orders of magnitude times larger than the Sloan Digital Sky Survey (SDSS) and the Galaxy And Mass Accretion (GAMA) survey combined). Combined with deep$+$high spatial resolution imaging from the HSC, this program will result in the first high S/N weak lensing profile measurements for dwarf galaxies, probing their DM halos out to their virial radii. Comparisons with state-of-the-art, detailed, hydrodynamical simulations will help to fully capture the interplay between DM, feedback, and their roles in the formation and evolution of dwarf galaxies. Merian is an important precursor to a much bigger program that can leverage the Rubin LSST. Custom narrow-band imaging over the LSST footprint affords the exciting opportunity to construct very large samples of dwarf galaxies and to measure their halo profiles with gravitational lensing. This program would provide novel and unique constraints on DM on dwarf scales. 

Even better constraints on the DM interaction cross section could be obtained by probing the central density of these dwarf halos in the region where a constant-density core could form in the presence of SIDM. This would require measurements of their central velocity dispersions, $\sigma_{*}$, which would necessitate a large spectroscopic survey of the Merian-found galaxies. This presents several challenges. First, dwarf galaxies ($M_{*} < 10^{9}M_{\odot}$) are intrinsically faint, making absorption-line dispersion measurements difficult.  Next, for the more distant galaxies, 8--10 m class telescopes are required, but fields-of-view and scarce observing time limits the final sample size. Smaller telescopes (2--4 m) offer more access with wider fields, but struggle to reach the required spectroscopic depths.

A way forward is offered by new astrophotonic ``spectrometer-on-a-chip'' technology coupled with coming wide-field adaptive optics (e.g., ground-layer AO) systems. Such systems are being planned for a variety of telescopes of modest size and should eventually provide a 0.8 deg$^{2}$ field-of-view with performance at $r$-band wavelengths. The AO systems offer sensitivity gains thanks to improved seeing.  In addition, photonic spectrometer arrays hold promise for enabling optimal extraction of turbulent modes, thereby offering further gains in S/N.  
If realized, these technologies would enable a 2--4 m telescope to achieve S/N $\approx$ 10 $\AA^{-1}$ at $r_{AB} \approx 21.5$, sufficient for $\sigma_{*}$ measurements to 20\%. An instrument with these capabilities and a 150 multiplex across 0.8 deg$^{2}$ could observe a 10,000 dwarf sample over $\sim$200
nights.  On a 2--4 m class telescope, such an experiment would, at a modest cost, deliver enormous value, with SIDM cross-section constraints improved by several factors to an order-of-magnitude over current constraints.

\subsubsection{Milky Way stellar dynamics}
\label{stellardynamics}

The gravitational potential of the Milky Way can be measured by studying the dynamics of stars. Such dynamical mass measurements are useful for constraining the local DM density, which is a crucial quantity for direct and some indirect detection experiments \citep{Jungman:1995df,2015PrPNP..85....1K}, as well constraining DM particle models that give rise to sub-structures or have a characteristic larger scale spatial density dependence. The astrometric data that is used for such measurements has improved tremendously in the last few years thanks to the \emph{Gaia} satellite \citep{2018A&A...616A...1G}, whose catalogue contains over a billion stars with parallax and proper motion information.  At the same time, advances in technology in the coming decade will enable ``real-time'' Galactic dynamics wherein extremely precise time-series measurements can be used to directly measure Galactic accelerations, and therefore the Galactic potential.

Dynamical mass measurements are typically performed under the assumption of a steady state as well as a few simplifying symmetries (e.g.~rotational symmetry or mirror symmetry with respect to the Galactic plane), often reducing the dynamics to only one or two spatial dimensions (e.g.~the rotational velocity curve or the vertical oscillation of disk stars). In studies that determine the local DM density, the suite of results from different \emph{Gaia}-based studies are somewhat discrepant. Moreover, where the results of local analyses (e.g. based on the vertical oscillations of stars within roughly kilo-parsec distance) are found in the range 0.4--0.6~GeV/cm$^3$, while the results of more global analysis (e.g.~based on the circular velocity curve) are found in a slightly lower range of 0.3--0.5~GeV/cm$^3$ \citep{2021RPPh...84j4901D}. The systematic source of these discrepancies could be a mismodeled distribution of baryons or incorrect assumptions of a steady state or symmetry \citep[e.g.][]{2019JCAP...10..037D,2020MNRAS.495.4828G,2021RPPh...84j4901D}. 

This suite of measurements does not only inform us of the local DM density, but can also constrain DM particle models. Analyses of the Milky Way's rotational velocity curve have been used to place constraints on superfluid dark matter, which can exhibit MOND-like acceleration profiles on smaller spatial scales \citep{2019arXiv191112365L,2020MNRAS.498.3484H}. Very local analyses of the vertical oscillation of stars have placed limits on a so-called thin dark disk, which would consist of a dark matter subcomponent with strong dissipative self-interactions \citep{Fan:2013tia,Fan:2013yva}. Current limits to a thin dark disk's surface density are roughly 5~M$_\odot$/pc$^2$ \citep{Schutz:2017tfp,Buch:2018qdr,2021A&A...653A..86W}, such that a thin dark disk is allowed only due to uncertainties associated with the baryonic matter density distribution (at least under the assumption that the thin dark disk is stable and co-planar with the stellar disk).

There is progress in relaxing these assumptions of a steady state and Galactic symmetries, which will be necessary to make further progress in extracting information about DM microphysics from galactic dynamics using \emph{Gaia} (noting again that \emph{Gaia} data is precisely what has allowed us to uncover the deviations from these assumptions). For example, the full three-dimensional gravitational potential can analysed with a method using ``normalising flows'' \citep{2021MNRAS.506.5721A}, which is model independent in terms of symmetry assumptions. It is also possible to relax the assumption of a steady state and extract information from the shape of time-varying dynamical structures of the Milky Way; this is exemplified by a new method \citep{2021A&A...650A.124W,2021A&A...653A..86W} that infers the gravitational potential using the recently discovered phase-space spiral \citep{2018Natur.561..360A}.

Beyond these modeling improvements, it has recently become possible to directly measure the very small Galactic accelerations of stars ($\sim$ 10 cm/s/decade) using time-series precision measurements with several different techniques. By measuring a direct acceleration, it is not necessary to assume anything is in equilibrium since accelerations are directly related to gradients of the potential via the Poisson equation. Recent work has shown that spectral measurements of the line-of-sight (radial) velocities (RVs) of individual stars will be sensitive to the local DM density~\cite{Quercellini:2008it,Ravi:2018vqd,Silverwood:2018qra,Chakrabarti2020}, for instance using the current generation of high-precision spectrographs like ESPRESSO \cite{Pepe2010} that have achieved precision RV precision $\sim$ 10 cm/s \cite{Wright_Robertson2017}. When extremely large telescopes like the Giant Magellan Telescope begin observations, extreme-precision radial velocity measurements can be carried out across the Galaxy~\cite{Gclef}. Line-of-sight accelerations
of pulsars measured via their spins~\cite{Phillips:2020xmf} and orbital periods~\cite{Phillips:2020xmf,Chakrabarti2021} may already have hints of the local Galactic
acceleration, with improvements possible in the future. It is also now imminently possible to measure the Galactic acceleration directly using eclipse timing, by measuring the small shift in the mid-point of the eclipse time over the past decade induced by the Galactic potential since the \textit{Kepler} mission \citep{ChakrabartiET}, using space-based observations from \textit{HST}, and in the future with \textit{JWST} and \textit{Roman} observations. \textit{Gaia} astrometric observations were also recently analyzed to measure the solar system acceleration \citep{Klioner2021,2020arXiv201202169B}, which is in agreement with prior measurements by VLBI \citep{Charlot2020}; given these measurements, it may be possible to use statistically aggregated astrometric accelerations of stars to measure the galactic potential~\cite{Buschmann:2021izy}, which would be aided further by astrometric missions beyond \emph{Gaia} like \emph{Theia}~\cite{Theia:2017xtk} or \emph{GaiaNIR}~\cite{2016arXiv160907325H}. Thus, there is now a wide range of techniques available currently or on the timescale of the next decade that will enable ``real-time'' Galactic dynamics. In the future, this will allow for precise characterizations of the mass distribution (both the smooth component and dark matter sub-structure) across the entire Galaxy, without assumptions of equilibrium or symmetry. These measurements can constrain DM physics due to the different spatial morphology and clustering of DM substructure in different models of DM as detailed in Section~\ref{sec:theory}.

\subsubsection{Galactic rotation curves}
\label{sec:rotation}

\begin{figure}[h]
\centering
\includegraphics[scale=0.36]{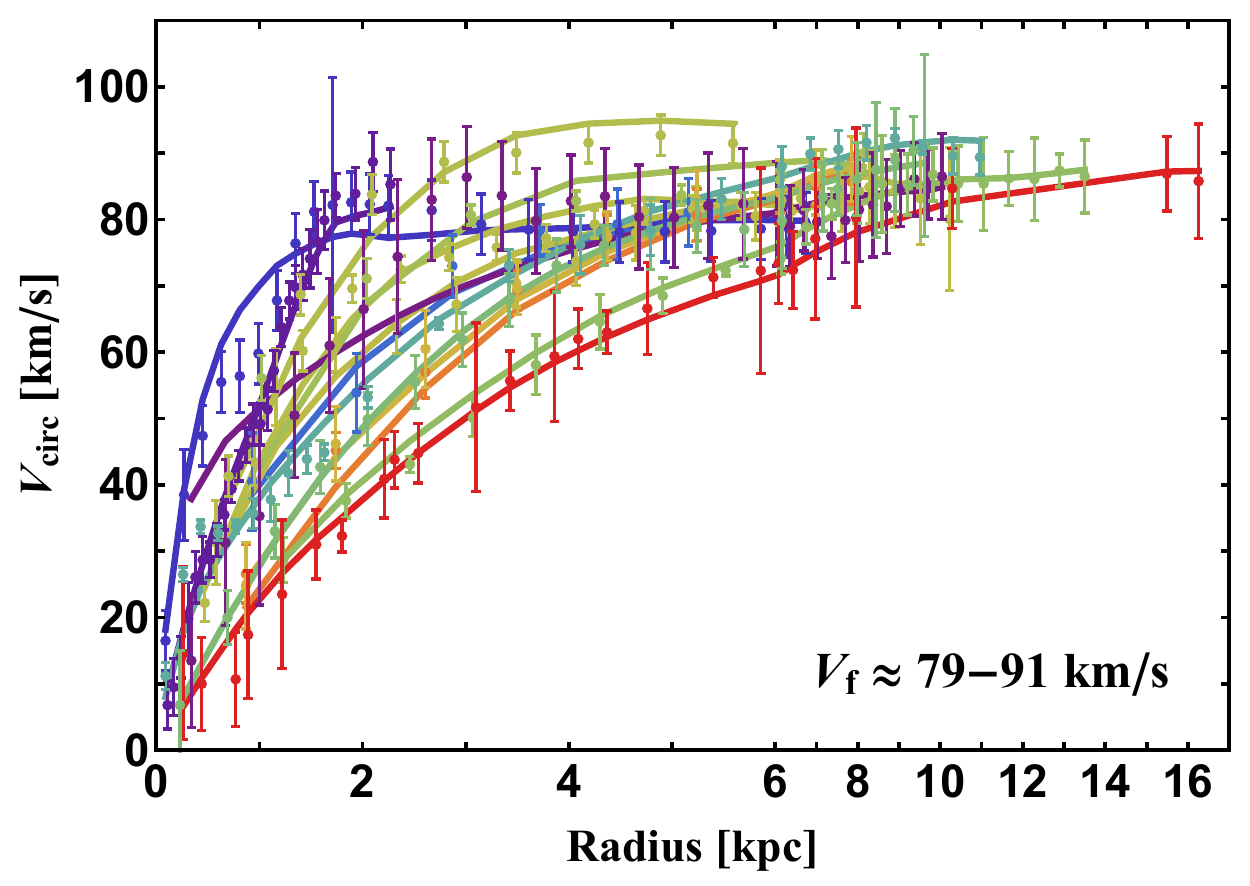}
\includegraphics[scale=0.37]{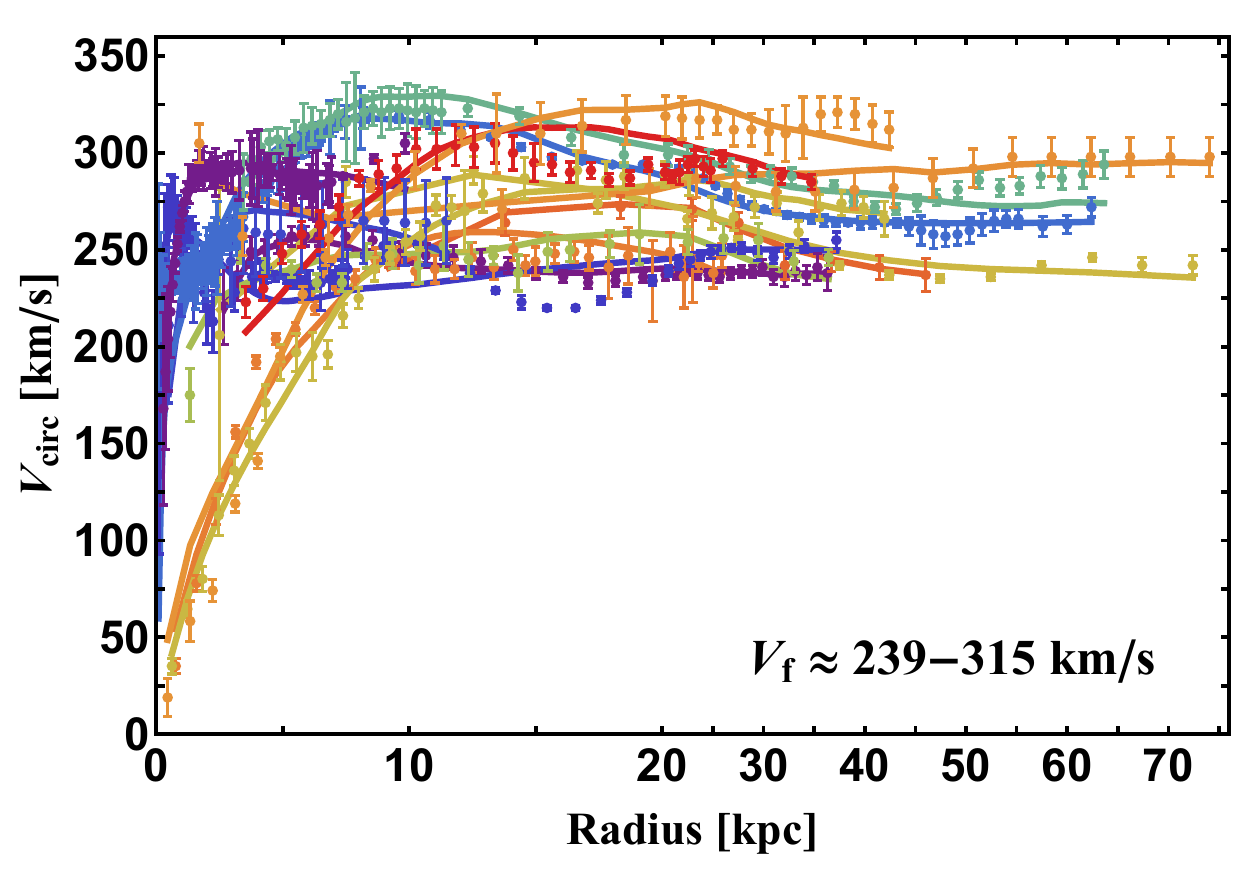}
\caption{Diverse galactic rotation curves of spiral galaxies. The data points with error bars are from the SPARC dataset~\cite{Lelli:2016zqa}. Each panel contains $14$ rotation curves that are selected to have similar asymptotical velocities at further radii. The velocity bins are $79\textup{--}91~{\rm km/s}$ (left panel) and $239\textup{--}315~{\rm km/s}$ (right panel), representing small and large mass limits of the sample, respectively. Each solid curve denotes the SIDM fit to an individual galaxy. Taken from Ref.~\cite{Ren:2018jpt}.\label{fig:diversity}}
\end{figure}

The CDM model is extremely successful in explaining the matter distribution of the Universe on large scales. However, it has challenges in matching with observations on galactic scales. Dwarf and low surface brightness galaxies often prefer a cored DM density profile rather than a cuspy one predicted in CDM-only simulations, which was initially coined as the ``core vs cusp" problem, see~\cite{deBlok:2009sp} for a review. However, galaxies are not uniformly cored~\cite{KuziodeNaray:2009oon}, and the inner rotation curves of spiral galaxies exhibit diverse shapes with a spread larger than predicted in CDM~\cite{Oman:2015xda}. More recently, the issue has been reframed as the ``diversity" problem. 

Figure \ref{fig:diversity} shows circular velocity profiles of spiral galaxies from the SPARC sample~\cite{Lelli:2016zqa}. The left panel contains galaxies with low masses and their asymptotical velocities are $V_{f}\approx 79\textup{--}91~{\rm km/s}$, while the right panel contains high-mass ones with $239\textup{--}315~{\rm km/s}$. For both cases, the spread in the circular velocity at $\sim2~{\rm kpc}$ is a factor of $\sim4$, although the galaxies in each mass bin have similar velocities at further radii.

In CDM, the feedback processes associated with star formation could alter the gravitational potential and change the DM distribution in the inner halo. Simulations show that if the feedback is driven by spatially and temporally fluctuating star formation, a cored CDM halo could be produced~\cite{Governato:2009bg,DiCintio:2013qxa,Chan:2015tna}, and hence the predicted rotation curve rises slowly. On the other hand, the halo remains cupsy if the feedback is driven by smoother star formation~\cite{Oman:2015xda,Fattahi:2016nld}. The challenge is to produce both cored and cuspy halos in simulations after adapting a specific feedback model. 

Recent studies show that the SIDM model~\cite{Spergel:1999mh,Kaplinghat:2015aga} can explain the full range of the diversity~\cite{Kamada:2016euw,Creasey:2017qxc,Ren:2018jpt,Kaplinghat:2019dhn}. The SIDM model predicts both cored and cuspy profiles, depending on baryon concentration~\cite{Kaplinghat:2013xca}. In DM dominated galaxies, thermalization due to the self-interactions creates large cores and reduces DM densities. In contrast, thermalization leads to denser and smaller cores in more luminous galaxies. Ref.~\cite{Ren:2018jpt} analyzed the rotation curves of $135$ galaxies from the SPARC dataset and found that SIDM provides excellent fits, as demonstrated in Fig.~\ref{fig:diversity} (solid curves). The inferred halo parameters are consistent with the concentration-mass relation from cosmological CDM simulations, indicating that SIDM and CDM predictions are consistent on large scales.

Looking forward, there are several promising directions that we can take. The analysis in Ref.~\cite{Ren:2018jpt} assumes a static stellar distribution from observations as an input. This approach is well-justified for the purpose of fitting to the rotation curves at $z=0$. However, to make further progress, we need to understand the formation history of the stellar distribution. In fact, a key challenge for structure formation simulations in both CDM and SIDM is to reproduce the large spread in the stellar density to be consistent with the observations~\cite{Kaplinghat:2019dhn}. To resolve this, simulations may need to adopt different baryonic feedback models depending on the DM properties. In addition, it is important to analyze the rotation-curve data in  other DM scenarios beyond CDM, such as warm DM and fuzzy DM. To perform an efficient statistical analysis, we need to develop a semi-analytical approach to model DM distributions with the presence of baryons for those models, as in the case of SIDM. 

Furthermore, recent observations have discovered a large number of ultra-diffuse galaxies that exhibit a much more extended baryon distribution~\cite{2015ApJ...798L..45V,2015ApJ...807L...2K,2019ApJ...883L..33M,2020NatAs...4..246G}, compared to ordinary spiral galaxies. Some of them have high baryon contents, but low circular velocities, indicating that they are baryon-dominated and their host halos must have extremely low concentrations~\cite{2021ApJ...909...20S,2021MNRAS.tmp.3311M}. In the near future, high-resolution measurements of stellar and gas kinematics will be available for more ultra-diffuse galaxies, and they will provide an exciting opportunity for testing different DM and feedback models. In the next decade, a collaborative effort among particle physics theorists, simulators and observers is essential for seeking a successful structure formation theory which could provide a  uniform explanation to observations of galactic systems on all scales.

\subsubsection{Ultraviolet luminosity function}

 Observations of galaxies at high redshifts provide us with a unique and highly complementary handle on the physics of structure formation during the epoch of cosmic reionization~\cite{finkelstein_2016_review}. 
 
\begin{wrapfigure}{R}{0.55\textwidth}
\includegraphics[width=\linewidth]{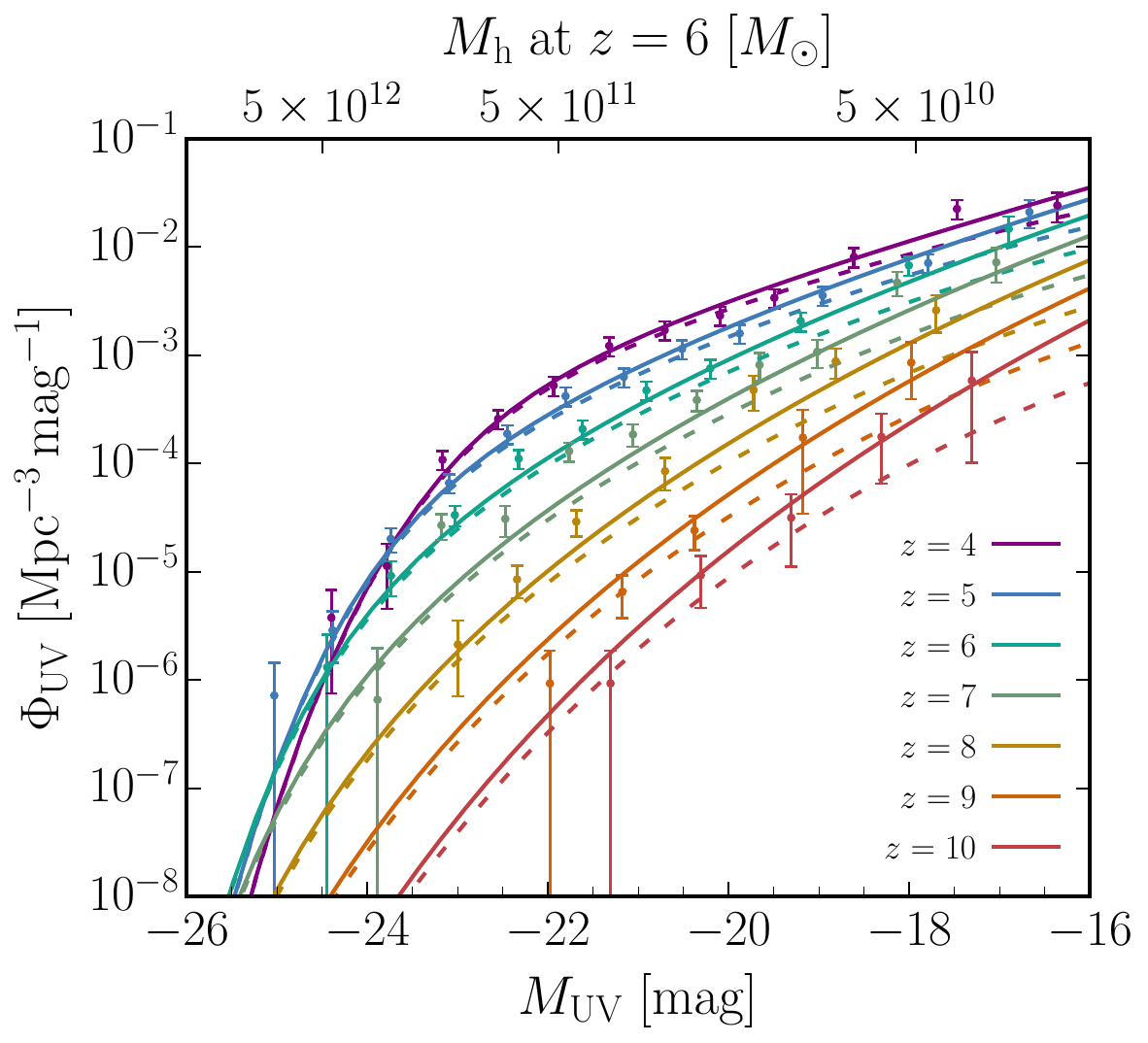}
\centering    
\caption{The UV LF as a function of absolute magnitude. Data are from HST observations~\cite{Oesch_2018, Bouwens_2021}, whereas models are from Ref.~\cite{Sabti:2021xvh}. Solid curves are for CDM and dashed are for 5 keV thermal WDM (assuming same astrophysics). The top axis shows the corresponding halo masses at $z = 6$. Figure adapted from Ref.~\cite{Sabti:2021unj}.}
\label{fig:UVLF1}
\end{wrapfigure}
\noindent The abundance of galaxies is measured by the Ultraviolet luminosity function (UV LF), and is controlled by both the astrophysics of galaxy formation and the clustering of DM. The galaxies forming during reionization $(z=4-10)$ have been mapped by the Hubble Space Telescope through a decades-long observational campaign that has resulted in two main catalogs --- the Hubble Legacy Fields (HLF) and Hubble Frontier Fields (HFF) --- that cover tens of thousands UV-bright galaxies, see e.g.~\cite{Bouwens:2014fua,Finkelstein_2015,Atek:2015axa,Livermore:2016mbs,Bouwens_2017asdasd,Mehta_2017,Ishigaki_2018,Oesch_2018,Atek:2018nsc,Rojas_Ruiz_2020, Bouwens_2021}. The measured UV LF from Refs.~\cite{Oesch_2018, Bouwens_2021} is shown in Fig.~\ref{fig:UVLF1}, together with the best-fit model from Ref.~\cite{Sabti:2021xvh}. The top axis shows the corresponding halo masses covered at $z=6$, defining the scales that we can access with this probe ($M_\mathrm{h} \sim 10^{10} - 10^{12}\, M_\odot$ or, equivalently, $0.1\,\mathrm{Mpc}^{-1} < k < 10\,\mathrm{Mpc}^{-1}$).

From a modeling perspective, several different approaches have been employed to predict the UV LF, ranging from (semi-)analytical methods~\cite{mashian2015empirical, Yung_2018, Park:2018ljd, Gillet:2019fjd}, to more elaborate techniques that are calibrated to simulations~\cite{Cai:2014fja, Sun_2016, Tacchella:2018qny, Behroozi:2019kql}, and full-scale hydrodynamical simulations~\cite{Salvaterra:2010nb, Jaacks, Dayal:2012ah,oshea2015probing, Vogelsberger_dust2020}. A crucial ingredient in constructing the UV LF is the so-called galaxy--halo connection~\cite{Wechsler:2018pic}, which encodes the relation between the UV magnitude of the galaxy and the mass of its host halo. The astrophysical parameters that enter this link can be highly degenerate with cosmological parameters, see e.g.~\cite{Rudakovskyi:2021jyf,Sabti:2021xvh} for an illustration, making it necessary to carefully model and marginalize over the unknown astrophysics.

The UV LF has seen many applications that improved our understanding of cosmology at high redshifts. In particular, previous works have exploited this probe to measure $\sigma_8$~\cite{Sahlen:2021bqt, Sabti:2021xvh}, and to derive constraints on warm- and fuzzy-DM models~\cite{Bozek:2014uqa, Schultz:2014eia, Dayal:2014nva, Corasaniti:2016epp, Menci:2017nsr, Menci:2018lis, Rudakovskyi:2021jyf}, dynamical dark energy~\cite{Menci:2020ybl}, the primordial matter power spectrum~\cite{Yoshiura:2020soa}, as well as the small-scale power spectrum~\cite{Sabti:2021unj} and non-Gaussianities~\cite{Chevallard:2014sxa, Sabti:2020ser} beyond the reach of the Comic Microwave Background.

Future advancements with the UV luminosity function will come from both the observational and modeling side. In the former case, upcoming observatories like the James Webb Space Telescope~\cite{Gardner:2006ky} and Roman Space Telescope~\cite{Spergel:2015sza} will cover larger areas and fainter magnitudes than currently possible. Measurements of the faint end of the UV LF will particularly be important for understanding the clustering properties of DM, as such observations cover smaller halo masses. Besides improved data, one of the main challenges will be to better understand the dependence of the cosmological results on the astrophysical modeling. This can be done by studying the galaxy--halo connection in hydrodynamical simulations in order to control the degeneracies between astrophysical and cosmological parameters. Insights from other probes, such as 21-cm maps (see Section \ref{subsec:21cm}), may also prove to be useful in this regard. 

\subsection{Theory Connections}
Detailed measurements of the DM density profile of luminous galaxies provide a window into the \emph{in situ} impact of DM physics within these objects. Such observations are for instance probing the SIDM cross section at typical galactic velocities (see Section \ref{sec:SIDM}), or the presence of a solitonic core caused by ultra-light bosonic DM (see Section \ref{sec:FDM}). Furthermore, galactic halo profile measurements are also sensitive to the presence of dissipation in the DM sector (which could happen in a composite DM scenario, see Section \ref{subsec:composite}) or even to halo energy injection through either DM annihilation or decay (see e.g.~Refs.~\cite{2019MNRAS.tmp....3R,Wadekar:2019mpc,2021arXiv211108025W}). On the other hand, the onset of galaxy formation in the earlier Universe as probed by the UV LF is very much influenced by \emph{ab initio} DM properties. This makes such observations sensitive to the DM thermal history (see Sections \ref{sec:WDM} and \ref{sec:non-therm}), as well as to any nongravitational interactions that DM particles experienced in the early Universe (see Sections \ref{subsec:bar_neut_scat} and \ref{subsec:composite}).

\section{Beyond the Threshold of Galaxy Formation} \label{sec:beyond_galaxy_threshold}

\subsection{Scale and Objects}

The formation of stars, and hence entire galaxies in the process, requires gas to efficiently cool to further collapse into ever denser regions in the vicinity of radiation coming from the first generation of stars.
The limit where atomic hydrogen can cool efficiently is at around virial temperatures of $T_{\rm vir} \sim 10^4 K$ which corresponds to DM halos at around $\sim 10^{7-8}M_{\odot}$. Current observations presented in this Section are approaching the limit to empirically measure the minimum halo mass where where stars were formed.
At least some halos in the mass range of $\sim 10^{8-9}M_{\odot}$ do host stars. These halos can be observed as ultrafaint galaxies. These systems are currently the most DM-dominated systems with a ratio of DM to stellar mass significantly exceeding a factor of a hundred.
DM halos below this threshold are expected to be depleted of stars and are effectively invisible. Halos at and below the galaxy formation threshold offer a pristine environment in which baryonic physics is subdominant to the dark matter behavior, and hence are substantially less affected by baryonic systematics.

\subsection{Observational Probes}

In the following we list some of the promising current and future probes that can probe the abundance and inner composition of halos in the mass range of $10^{6-9}M_{\odot}$.

\subsubsection{Ultra-faint galaxies}

Ultra-faint dwarf (UFD) galaxies, defined as dark matter-dominated stellar systems with luminosity $L < 10^5\ L_{\mathrm{\odot}}$ \citep{Simon190105465}, occupy the smallest dark matter halos that form galaxies, with $10^{8}\ M_{\mathrm{\odot}} \lesssim M_{\mathrm{halo}}\lesssim 10^{10}\ M_{\mathrm{\odot}}$ \citep{Jethwa161207834,DES:2019ltu}. These systems therefore provide crucial information about the astrophysical processes that govern the faint end of galaxy formation and the dark matter physics that impacts the abundance and properties of halos near the galaxy formation threshold.

To date, ultra-faints have exclusively been detected within distances of $\sim 300\ \mathrm{kpc}$ as satellites of the Milky Way. Deep photometric surveys including the Dark Energy Survey have significantly increased the number of known Milky Way satellites in recent years, which now exceeds $\sim 60$ systems \citep{Drlica-Wagner191203302}. Nevertheless, both empirical and theoretical arguments suggest that our census of ultra-faints---even within the Milky Way’s virial radius---is highly incomplete \citep{Tollerud:2008ze,Hargis:2014kaa,Newton170804247}. Forthcoming observational facilities including the Vera C.\ Rubin Observatory and the Nancy Grace Roman Space Telescope are expected to drastically improve upon current ultra-faint detection capabilities. Specifically, Rubin is forecasted to discover the entire population of UFD Milky Way satellites in the Southern hemisphere outside the Galactic disk and with surface brightness down to $\sim 32 \mathrm{mag\ arcsec}^{-2}$ \citep{LSST:2008ijt,LSSTDarkMatterGroup:2019mwo}. Excitingly, these detections are expected to extend well beyond distances of ~300 kpc, allowing UFD populations to be characterized throughout much of the Local Volume \citep{Mutlu-Pakdil210501658}.

Our understanding of the dark matter halos that host ultra-faints has significantly improved over the last decade. In particular, hydrodynamical simulations have started to resolve the formation and evolution of UFDs in a cosmological context, including as satellites of larger hosts \citep{Wheeler150402466,Wheeler181202749,Munshi181012417,Agertz190402723,Applebaum200811207,Grand210504560}. In parallel, semi-analytic and empirical galaxy--halo connection models have been combined with observations to infer ultra-faints’ halo properties; in these models, the number of observable ultra-faints suggests that these systems occupy halos with masses down to $\sim 10^8\ M_{\mathrm{\odot}}$ (near the ``atomic cooling limit''; \cite{Jethwa161207834,Kim:2017iwr,Nadler:2018iux,DES:2019ltu}) or even lower \citep{Kelley:2018pdy,Graus180803654}. The fraction of low-mass halos that host UFDs and the scatter in the stellar mass--halo mass relation for these systems are also being studied in order to precisely characterize the properties of halos near the galaxy formation threshold \citep{Rey:2019zsq,Benitez-Llambay200406124,Munshi210105822,Santos-Santos211101158}.

Ultra-faints provide stringent constraints on dark matter particle properties. Specifically, the number of observable UFD Milky Way satellites informs dark matter physics that suppresses the present-day abundance of dark matter halos with masses of $\sim 10^8\ M_{\mathrm{\odot}}$, yielding lower limits on the mass of thermal relic warm dark matter of $\sim 5\ \mathrm{keV}$ \citep{Maccio09102460,Polisensky:2010rw,Anderhalden12122967,Lovell:2013ola,Kennedy:2013uta,Schneider:2014rda,Cherry:2017dwu,Kim:2017iwr,DES:2020fxi,Newton:2020cog,Dekker:2021scf} and on the mass of ultra-light dark matter of $\sim 10^{-21}\ \mathrm{eV}$ \cite{Safarzadeh:2019sre,DES:2020fxi}. Similar analyses constrain the strength of dark matter--Standard Model interactions at the $\sim 10^{-30} \mathrm{cm}^2$ level \citep{Nadler:2019zrb,DES:2020fxi,Maamari:2020aqz,Nguyen:2021cnb} and the dark matter particle lifetime at the $\sim 10\ \mathrm{Gyr}$ level \citep{Peter:2010sz,Wang:2014ina}. Furthermore, stellar velocity dispersion measurements for brighter dwarf galaxies inform the strength and velocity-dependence of dark matter self-interactions, with sensitivity to cross sections of $\sim 1 \mathrm{cm}^2\ \mathrm{g}^{-1}$ \citep{Read:2018pft,Correa200702958}, and combining measurements of UFD internal dynamics from upcoming spectroscopic facilities and giant segmented mirror telescopes \citep{Simon:2019ojy} with their population statistics will further inform self-interacting dark matter models \citep{Kim210609050,Nadler:2020ulu,Nadler:2021rpo}.

The forthcoming observations discussed above are expected to increase the sample of known UFDs near the galaxy formation threshold, reducing statistical uncertainties and improving the corresponding dark matter constraints. For example, Rubin observations are forecasted to probe halo masses below $\sim 10^8\ M_{\mathrm{\odot}}$ \citep{LSSTDarkMatterGroup:2019mwo}. If these halos are occupied by ultra-faints, these observations will provide evidence for the existence of galaxies formed through molecular hydrogen cooling; otherwise, they will determine the halo mass threshold and stochasticity associated with galaxy formation. Thus, combining these measurements with purely gravitational probes of small-scale structure will enable the detection of completely dark halos.

\subsubsection{Lyman-alpha forest}

The intergalactic medium (IGM) is the rarefied gas that fills the vast volumes between the
galaxies in the Universe. Physical effects, ranging from the nature of dark matter to the radiation from star-forming galaxies and quasars, set the observable properties of the
IGM, making the IGM a powerful probe of both fundamental physics and astrophysics.
Baryons in the IGM trace dark matter fluctuations on Mpc scales, while on smaller ($\lesssim$100 kpc scales) the $T\sim10^4$K gas is pressure supported against gravitational collapse. While the IGM is sensitive to the Epoch of Reionization \citep{Onorbe2017,Onorbe2017b,boera19}, complex and poorly understood physical processes related to galaxy formation play only a minor role in determining its structure \cite{Kollmeier2006, Desjacques2006}.

Forward modeling the structure of the IGM for a given cosmological and DM scenario is a theoretically well-posed problem, albeit requiring expensive cosmological hydro-dynamical simulations. Despite this apparent simplicity, accurate simulations of the IGM are computationally challenging because of the multi-scale nature of the problem \cite{Lukic2015}. A spatial resolution of $\approx 20~{\rm kpc}$ is required to resolve the density structure of the IGM, while a relatively large box size is required to capture the large-scale power, to obtain a fair sample of the Universe, and to model the topology of reionization.

These characteristics of the Lyman-$\alpha$ flux power spectrum have allowed for some of the tightest constraints on the nature of dark matter clustering \citep{irsic17b,yeche17,palanque19,irsic17c,armengaud17,rogers20}. Current analyses have been shown to be sensitive to the scale of either a suppression \citep{murgia17,murgia18} or an enhancement\citep{murgia19,irsic19} of the small scale power. These dark matter constraints come from high-quality observations of a the few currently available ($\sim 100$) high redshift ($z>4$) quasar spectra obtained with high resolution instruments ($R\sim 80,000$) and high signal-to-noise observations ($S/N\sim 100$). With these spectra, surveys are able to measure scales to wavenumbers of $k \sim 20\;\mathrm{h/Mpc}$ \citep{walther18,boera19}.
Notably these surveys are currently limited by statistical errors.

Analysis of the current data make clear that dark matter constraints benefit from access to higher redshifts ($z>4$). By improving both the theoretical predictions in this regime, as well as increased statistical power, we could push the lower bounds on the dark matter free-streaming scale by several orders of magnitude. Current bounds on the particle mass of various dark matter models include only a few spectra ($\sim 10-15$) at the highest redshifts ($z > 4.5$) (e.g. \citep{viel13wdm,irsic17a,rogers20}). Current or upcoming surveys will push this number to $\sim 25$ (DESI), $\sim 50$ (SkyMapper) and $\sim 100$ (LSST) at a fixed magnitude limit of $m_g < 18.5$. Each object requires $1$ night per quasar, so a survey of $3.5$ years can obtain $1000$ quasar spectra.

Current lower mass bounds for the WDM model are $m_{\rm WDM} > 2.1\;\mathrm{keV}$ using only high-redshift data and the most simplistic and uninformative priors on the thermal history evolution \citep{irsic17b}. Including lower redshift data pushes this bound to $> 3.5\;\mathrm{keV}$, and further including informative thermal history priors leads to $>5.1\;\mathrm{keV}$. Using a Fisher forecast analysis, at $1000$ spectra, the bound is $>14.4\;\mathrm{keV}$; almost an order of magnitude improvement over the current constraints coming from the equivalent high redshift data set analysis. This constraint maps to a lower bound on the ultra-light axion-like particle mass on the order of $\sim 10^{-21} - 10^{-20}\;\mathrm{eV}$. Combined with higher mass constraints from super-massive black hole super-radiance, this could exclude all ultra-light axions of mass $< 10^{-14}\;\mathrm{eV}$ as the dominant source of dark matter.

\subsubsection{Stellar streams}

Stellar streams, the tidal remnants of globular clusters and dwarf galaxies, are sensitive tracers of the distribution of matter at a range of scales throughout the Galaxy. At larger scales, stream morphology is sensitive to the global distribution of matter in the Milky Way, with recent studies of streams enabling constraints on the total mass and shape of the Milky Way halo and its massive satellites \citep{Erkal:2019, Vasiliev:2021, Shipp:2021}. At smaller scales, dynamically-cold stellar streams, resulting primarily from the disruption of globular clusters, retain information about interactions with small-scale perturbers \citep{Johnston:2016}, thus enabling tests of the clustering of dark matter at mass scales below the threshold of galaxy formation. Precise observation and modeling of a large population of stellar streams will therefore enable exciting constraints on the minimum dark matter halo mass, as well as measurements of the population of low-mass substructure in the Milky Way halo. 

In the simplest case, an interaction between a stellar stream and a dark matter subhalo leads to the formation of a gap in the distribution of stars along the stream. The depth and size of this feature can be used to infer the properties of the perturbing subhalo and the time since interaction \citep{Erkal:2015}. Many Milky Way streams have already been examined for gaps caused by interactions with low-mass subhalos, but at current observational limits it is difficult to unambiguously associate stream density variations with perturbations from subhalos. However, for strong encounters between a subhalo perturber and a thin stream, other features may appear in observed streams that provide even stronger (and less ambiguous) constraints on the properties of the perturber and its orbit. For example, thanks to recent data releases from the Gaia mission, these types of features were recently observed in two previously-known streams: A “spur” of stars has been found emanating from a prominent gap in the GD-1 stream \citep{Price-Whelan:2018} that has led to exciting constraints on the properties of the implied perturber \citep{Bonaca:2019, Bonaca:2020}, and more recently two streams have been kinematically associated into a single stream with a prominent “break” that may have been caused by another encounter from a different massive perturber \citep{Li:2021a}.

Studies of gaps in stellar streams provide insight into recent individual subhalo interactions. However, stellar streams are expected to experience many impacts over their lifetimes. Strong interactions produce clear gaps, while weaker interactions will lead to subtle variations in stream tracks and densities. All features will be washed out over time due to the internal velocity dispersion of stream member stars. The cumulative effect of a lifetime of perturbations can be studied statistically by measuring the power spectrum of density fluctuations along a stream \citep{Banik:2018, Bovy:2017}. Perturbations by dark matter subhalos at various masses, as well as by other Milky Way structures, result in density fluctuations at different scales. Measurements of the density power spectrum can therefore reveal the interaction history of a stellar stream, and the mass distribution of the perturbing subhalo population. Density power spectra have been used to provide evidence for a population of dark subhalos around the Milky Way and constrain a range of dark matter models \citep{Banik:2021}. Future studies with deeper observations and larger population of streams are projected to constrain the subhalo population down to $10^5 M_\odot$ \citep{Bovy:2017}.

Recent large photometric and astrometric surveys have enabled the discovery of more than 60 streams around the Milky Way \citep{Grillmair:2016, Shipp:2018, Malhan:2018}. Upcoming surveys, such as the Rubin Observatory Legacy Survey of Space and Time \citep[LSST;][]{Ivezic:2019} will enable the discovery of streams at larger distances and lower stellar densities, and will improve observations of known streams, reaching fainter member stars and obtaining higher photometric precision. LSST is predicted to enable detections of gaps caused by subhalos with masses down to $10^6$ to $10^7 M_\odot$ for streams with a range of typical surface brightnesses at a distance of 20 kpc \citep{Drlica-Wagner:2019}. 

Precise kinematic measurements are also necessary to accurately infer the perturbation history of stellar streams. Kinematic measurements facilitate the separation of stream members from foreground contamination, and enable the dynamical modeling of these systems. The sensitivity of streams to baryonic perturbers varies based on the stream orbit. Precise orbit modeling is therefore necessary to select a clean sample of streams and to model any extraneous sources of perturbation. The Gaia Mission has provided unprecedented proper motion measurements of more than 1 billion stars in our Galaxy \citep{Gaia:2018}, and has enabled measurements of the proper motions of Milky Way streams out to $>50$ kpc \citep{Shipp:2019}. Radial velocities have historically been difficult to measure due to the diffuse, extended nature of these systems. However, by taking advantage of cutting edge photometric and astrometric datasets to efficiently select targets for observation, the Southern Stellar Stream Spectroscopic Survey \citep[S\textsuperscript{5};][]{Li:2019} has recently provided the first systematic spectroscopic observations of stellar streams, measuring radial velocities and metallicities of $>20$ stellar streams in the southern hemisphere.

These current and upcoming observations will enable detailed studies of individual perturbations and statistical measurements of the population of low-mass dark matter subhalos around the Milky Way. We can make predictions for the expected number of observable gaps in a given dataset within a particular dark matter mode \citep{Erkal:2016b}. For example, with LSST, within $\Lambda$CDM, we expect to observe 17 gaps in the 13 DES stellar streams, and a detection of fewer than 6 gaps would be inconsistent with $\Lambda$CDM at a $99.9\%$ level \citep{Drlica-Wagner:2019}.

\subsubsection{Galaxy strong lensing}
Strong gravitational lensing provides a unique probe of dark matter structure at the smallest cosmological scales, and thus constrains the particle nature of dark matter itself. By coupling directly to gravity, lensing circumvents luminous tracers of the underlying dark matter, and it is sensitive to mass scales below those of typical dwarf galaxies. Analyses of resolved distortions in extended Einstein rings of galaxy-scale strong lenses have resulted in the detection of substructure and halos along the line of sight with inferred lensing masses of $M_{\rm lens} \approx 10^{8-9}M_{\odot}$ with the Hubble Space Telescope, Keck adaptive optics (AO), and ALMA interferometric data \cite{Vegetti_2010_2,Vegetti_2012,Vegetti:2014,Hezaveh_2016_2,Birrer2017, Ritondale:2019,Sengul:2021lxe}.
The sensitivity of a lensing detection is limited by the angular resolution of the data, and by the structure of the lensed source. Very Long Baseline Array (VLBA) radio observations have the capacity to probe structure power at the $10^7 M_{\odot}$ scale, albeit for a very small sample of lens systems, and require analysis pipelines that can cope with the large data volume per observational target. 

The magnification ratios between adjacent images of multiply-imaged quasars provide an alternate means of probing substructure. Initial studies showed that these data are consistent with CDM at scales of $\gtrsim 10^{9}$~M$_{\odot}$ \cite{DK2002,Nierenberg:2017}. Recent progress with newly gathered data and a re-analysis of previously existing data resulted in tighter constraints demonstrating consistency with CDM to scales of $\gtrsim 10^{8}$~M$_{\odot}$ \cite{Gilman2020_wdm, Hsueh2020_wdm}. The latest constraints correspond to a free-streaming length scale of $\sim 10 \rm{kpc}$, or an equivalent thermal relic particle mass of $m_{\rm th} > 5$ keV, ruling out warmer mass functions at 2 sigma confidence. A similar study constraining the mass-concentration relation of dark matter halos at $\sim 10^{8}$~M$_{\odot}$ \cite{Gilman2020_mc} showed consistency with CDM predictions, and the extrapolation of N-body simulation results to the scales probed by lensing. In the context of self-interacting dark matter, mid-IR flux ratios obtained from the James Webb Space telescope will provide enough sensitive to detect populations of core-collapsed subhalos, probing the self-interaction cross section on velocity scales below $30 \ \rm{km} \ \rm{s^{-1}}$ \cite{Gilman2021_SIDM}. Constraints on halo mass function and concentration-mass relation from strong lensing analyses can also be interpreted in terms of the primordial matter power spectrum on scales $k > 10 \  \rm{Mpc^{-1}}$, irrespective of a particular dark matter model \cite{Gilman2021_Pk}. 

Conclusive results at the $10^7 M_{\odot}$ scale will require an expanded sample size of suitable lens systems, and improved data quality. The Rubin and Roman Observatories will yield up to 10,000 gravitational lenses, including several hundred quadruply-imaged quasars. The order of magnitude expansion of the current strong lens sample size forms the basis upon which strong gravitational lensing probes of dark matter will operate in the next decade. 

However, this sample of lenses will not translate into constraints on dark matter at the $10^7M_{\odot}$ level on its own; sufficiently precise and sensitive follow-up efforts must be undertaken for a targeted subsample. Gravitational imaging sensitivity will significantly benefit from the improved imaging resolution afforded by next-generation telescopes. The expected number of quadruply-lensed quasars on the sky is sufficient to meet the requirements to measure the mass structure at $\sim 10^7$~M$_\odot$ \cite{Gilman++19}, but the constraining power of a fixed sample size depends strongly on the measurement precision of the flux ratios. Current state-of-the-art measurements presented \cite{Nierenberg++20} achieve 5-6$\%$ precision. Considering that most of the quadruply-lensed quasars to be added to the existing sample will be fainter, enhanced observational capabilities will ensure that lensing reaches the $10^7 M_{\odot}$ target threshold. 

The James Webb Space Telescope (JWST) will enable precise spectroscopic measurement of narrow-line and dusty-tori flux ratios, but may not have the capacity to do so for a required sample of $\sim 100$ quadruply-lensed quasars. In addition, uncertainties in the emitting source structure affect the interpretation of the flux ratios and thus provide a systematic floor unless resolved measurements of a subset of quasars can be achieved.
The collecting area of the next generation of ground-based extremely large telescopes (ELTs; US-ELTP for the US has the strongest recommendation of the US Astro2020 Decadal Survey) with adaptive optics will make it possible to precisely measure flux ratios for a sufficiently large sample and allow a statistically precise inference of the structure at $\sim 10^7$~M$_\odot$.
Additionally, in order to fully exploit the future data, new methods have been developed to accelerate the detection of such systems as well as perturbers in the images \cite{DiazRivero:2019hxf,Alexander:2019puy,Varma:2020kbq, Ostdiek:2020cqz, Ostdiek:2020mvo} and to analyze the collective perturbation of a population of subhalos \cite{Hezaveh:2014aoa,DiazRivero:2017xkd,Daylan:2017kfh, Birrer2017, DiazRivero:2018oxk,Brennan:2018jhq,Brehmer:2019jyt}.

In conclusion, exploiting the full power of galaxy-scale strong lensing to constrain the nature of dark matter requires both exquisite data and sophisticated analysis tools. While the latter are being developed on precursor datasets, the former can only be obtained with the resolution and collecting area of extremely large telescopes (ELTs) or next generation VLBI. The analysis of the proposed data will require continued support for the development of methods, computational techniques and resources, as well as a robust interface between strong lensing predictions and cosmological numerical simulations of different dark matter theories.

\subsubsection{21-cm at cosmic dawn}\label{subsec:21cm}

The first galaxies that populated our Universe were hosted in dark-matter (DM) halos much smaller in size than their counterparts today, with masses in the range of $M_h\sim 10^{5-8}\,M_\odot$.
As a consequence, their abundance can be used to study matter fluctuations with very large wavenumbers ($k\sim 10-100\,\rm Mpc^{-1}$).
The 21-cm line of hydrogen maps the formation of these galaxies during cosmic dawn, at $z\sim10--20$, and thus provides a trove of information about the small-scale structure of dark matter.

The physics by which 21-cm photons are absorbed ~\cite{Furlanetto:2006jb,Pritchard:2011xb} from the CMB is as follows.
The UV light emitted by the first galaxies couples
the hyperfine degrees of freedom of hydrogen to
its thermal state through the Wouthuysen-Field (WF)
effect~\cite{Hirata:2005mz}.
This allows hydrogen to resonantly absorb
CMB photons with a 21-cm wavelength, 
which gives rise to 21-cm absorption (or a negative 21-cm
temperature brightness). 
Remnants from the first stellar formation (likely by X-ray binaries~\cite{HERA:2021noe}) will later heat up the hydrogen, producing observable 21-cm emission and ending cosmic dawn.

\begin{wrapfigure}{R}{0.55\textwidth}
\includegraphics[width=0.45\textwidth]{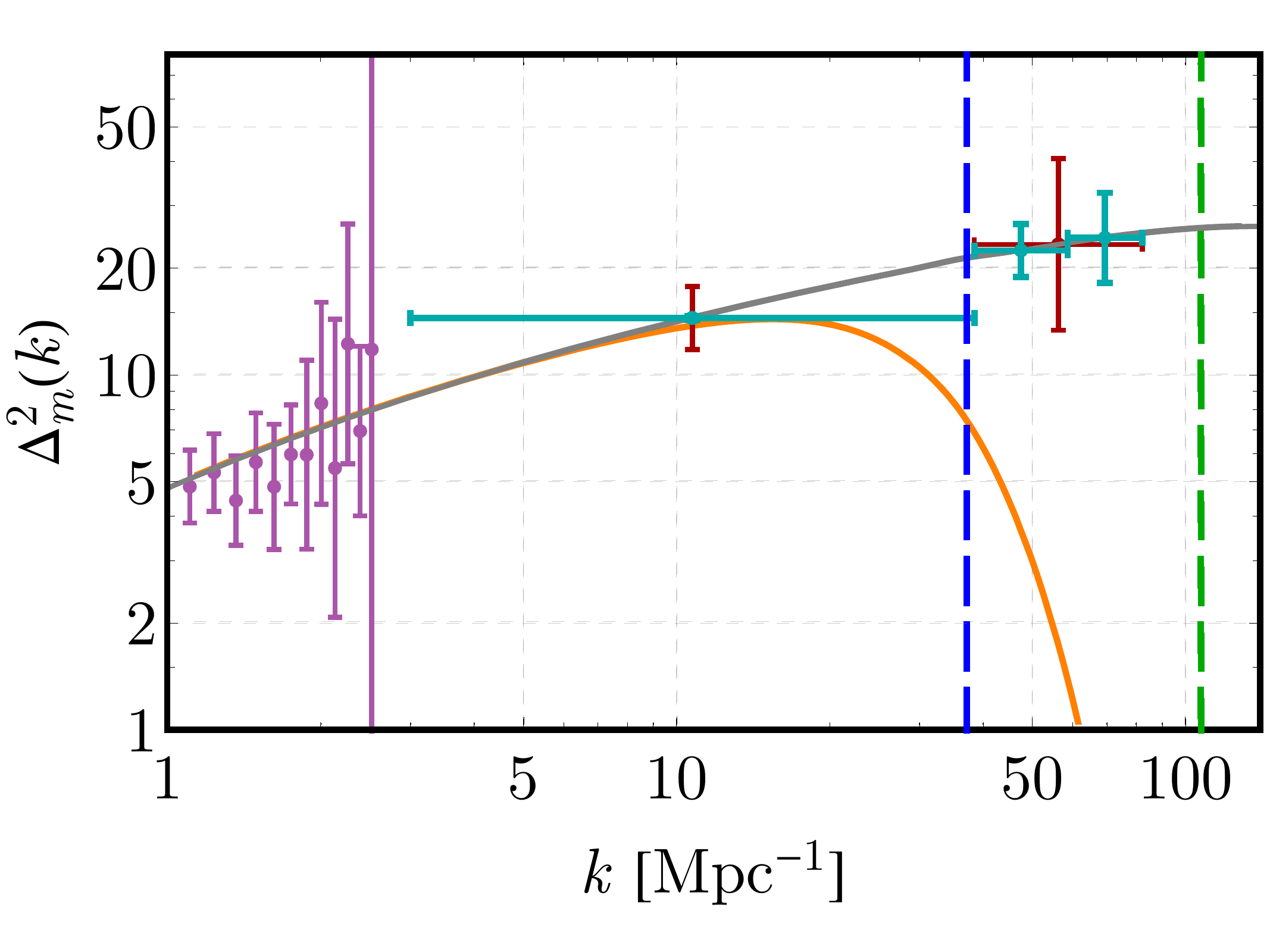}
\centering
\caption{Reach of 21-cm measurements on the small-scale matter power spectrum, linearly extrapolated to $z=0$. 
The crosses show the forecasts for an EDGES-{\it like} (red, global-signal) and HERA-{\it like} (cyan, fluctuation) experiment, which probe smaller scales than current public Lyman-$\alpha$ forest measurements (in purple).
The vertical lines correspond to the typical wavenumber that forms atomic-cooling (blue) or molecular-cooling (green) galaxies, which hosted the first stars.
The gray line corresponds to the standard CDM model, whereas orange represents WDM with $m_{\rm WDM}=5$ keV. 
Figure adapted from Ref.~\cite{Munoz:2019hjh}.}
\label{fig:21cm_forecast}
\end{wrapfigure}

The timing and shape of the 21-cm signal can be used to time the formation of the first galaxies, and thus to test the small-scale structure of DM on halos with $M_h\sim 10^{5-8}\,M_\odot$.
Models of non-cold DM that produce a suppression of power at small scales (e.g., warm, fuzzy, or interacting DM) will show a delay in the formation of the first galaxies, and thus of the 21-cm signal.
Likewise, an enhancement of power would appear as an earlier 21-cm cosmic-dawn trough, which can also be tested.

Cosmic-dawn 21-cm data comes in two flavors.
The first are measurements of the sky-averaged global signal (i.e., the monopole), which are experimentally less costly, requiring only one antenna, but more prone to systematics (e.g.,~\cite{Hills:2018vyr}). 
The first detection by the EDGES collaboration~\cite{Bowman:2018yin} (see, however, Ref.~\cite{Singh:2021mxo} which disfavors it) would indicate the presence of the first galaxies by $z=20$, when the signal goes into absorption.
If interpreted to be cosmological, this signal can be used to place limits on DM at small scales. 
For instance, the mass of warm DM (WDM) is bound to be above $m_{\rm WDM}>6.1$ keV~\cite{Schneider:2018xba} (see also~\cite{Safarzadeh:2018hhg,Boyarsky:2019fgp}), whereas for fuzzy (i.e., wave-like) DM its mass has to be above
$5\times 10^{-21}$ eV~\cite{Lidz:2018fqo}.
The EDGES signal can also also be used to measure the running of the scalar tilt at the 0.01 level~\cite{Yoshiura:2018zts}.
Note that this argument is independent of the anomalous EDGES depth~\cite{Munoz:2018pzp,Barkana:2018qrx,Berlin:2018sjs,Slatyer:2018aqg,Ewall-Wice:2018bzf}, as it relies just on the timing of the signal and not its amplitude.

The second are measurements of the 21-cm fluctuations, which are performed by interferometers such as HERA~\cite{DeBoer:2016tnn}, LOFAR~\cite{vanHaarlem:2013dsa}, or the SKA~\cite{Koopmans:2015sua}.
While these data require more antennas, they are expected to be cleaner of both systematics and foregrounds that plague the global signal~\cite{Liu:2009qga,Datta:2010pk,Parsons:2012qh}.
Current data just includes upper limits, and thus cannot be used to place limits on the small-scale structure of DM.
However, upcoming data on fluctuations will yield a clear picture of structure and galaxy formation at high redshifts~\cite{Mebane:2020jwl,Munoz:2021psm}.
For instance, HERA data will allow for a measurement of the small-scale matter power spectrum up to scales $k\approx 100\,\rm Mpc^{-1}$~\cite{Munoz:2019hjh}, see Fig.~\ref{fig:21cm_forecast}.
As a consequence, it will probe fuzzy or self-interacting DM beyond the reach of the global signal~\cite{Jones:2021mrs,Munoz:2020mue}.

In summary, the different astrophysical environment of the high-$z$ cosmic-dawn epoch, when the first galaxies formed, makes it an attractive target for understanding dark matter in a regime that is complementary to lower-$z$ observables.

\subsubsection{Extreme CMB lensing}

\begin{wrapfigure}{R}{0.5\textwidth}
\includegraphics[width=0.5\textwidth]{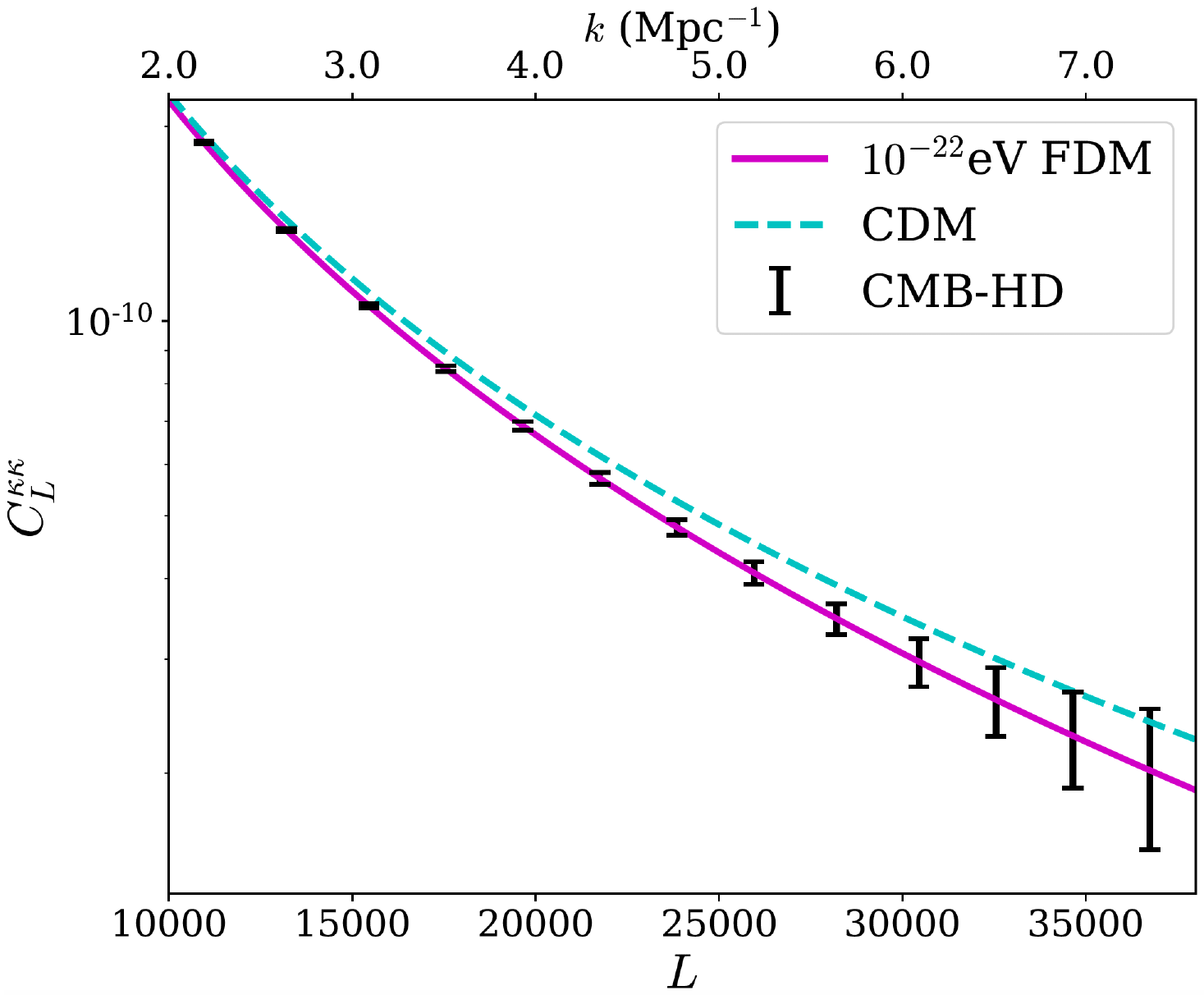}
\centering
\caption{Forecast for a CMB survey with 0.5 $\mu$K-arcmin noise in temperature and 15 arcsecond resolution over $50\%$ of the sky~\cite{Sehgal:2019ewc}. Such a survey would generate a high-resolution map of the projected matter distribution out to $k\sim10~h$Mpc$^{-1}$ over half the sky via gravitational lensing of the CMB.}
\label{fig:CMBLens}
\end{wrapfigure}

The gravitational deflection of the light of the Cosmic Microwave Background (CMB) by intervening structure has been demonstrated to be a powerful way to measure the distribution of matter on large-scales~\cite{Planck:2018lbu}, as well as the mass of halos with $M \ge 10^{13}$~M$_\odot$~\cite{ACT:2014ivk}. Measurements of the weak lensing of the CMB have the potential to probe the distribution of matter down to much smaller scales, given a CMB facility with sufficient sensitivity and resolution~\cite{Nguyen:2017zqu}.  Using gravitational lensing to probe the matter distribution directly avoids potential uncertainty in the use of baryons to trace the location of dark matter.  In addition, using the CMB as a backlight to measure gravitational lensing has the advantage that the CMB has 1)~a well-known redshift, 2)~well-known intrinsic statistical properties, and 3)~is behind all structure out to $z\sim 1100$. Moreover, the CMB lensing power spectrum is most sensitive to structure at $z\sim 2$; since structure builds up hierarchically that makes it more sensitive to models that suppress small-scale structure than a comparable lower redshift probe.

A CMB lensing measurement that could probe $k\sim10~h$Mpc$^{-1}$ and distinguish between the matter power spectrum predicted by CDM and that predicted by models that can explain observational puzzles of small-scale structure, with at least $5\sigma$ significance, requires instrument noise levels in temperature of 0.5 $\mu$K-arcmin and a resolution of $\sim15$ arcseconds over half the sky~\cite{Han:2021vtm,Sehgal:2019nmk,Sehgal:2019ewc,Nguyen:2017zqu} (see Figure~\ref{fig:CMBLens}). Such a measurement, which could be achieved by the proposed CMB-HD experiment~\cite{Sehgal:2019ewc,Sehgal:2020yja,CMB-HD-Snowmass}, would result in a high-resolution map of the projected dark matter distribution over half the sky. The CMB-HD survey would be three-times more sensitive and have six times higher resolution than CMB-S4; for comparison, CMB-S4 and precursor CMB experiments can reach $k\sim1~h$Mpc$^{-1}$ given their sensitivity and noise levels.

One challenge facing the interpretation of the small-scale dark matter distribution is the degeneracy between suppression of structure due to baryonic effects and suppression due to alternate models of dark matter.  However, the way baryonic effects and dark matter models impact the shape of the CMB lensing power spectrum is sufficiently different that they can potentially be differentiated~\cite{Nguyen:2017zqu}.  Another challenge is that such small-scale CMB lensing measurements at 0.5 $\mu$K-arcmin noise levels gain most of their signal-to-noise ratio from the lensing of CMB temperature maps, as opposed to polarization maps. However, CMB temperature maps have more foreground signals in them, such as the thermal Sunyaev-Zel'dovich (tSZ) effect, the kinetic Sunyaev-Zel'dovich (kSZ) effect, and the Cosmic Infrared Background (CIB).  Folding in the tSZ and CIB foregrounds to forecast what a CMB-HD survey could achieve results in distinguishing a benchmark $10^{-22}$ eV FDM model from a CDM model at the $5\sigma$ level~\cite{Han:2021vtm}.  Future work, with high-resolution millimeter-wave simulations and machine learning techniques are warranted to explore the full potential of foreground mitigation techniques.  For even lower noise levels than discussed above, the small-scale CMB lensing signal gains most of its signal-to-noise ratio from CMB polarization maps, which are much cleaner in terms of contaminating foregrounds.

\subsection{Theory Connections}

The experiments presented in this section will decisively test the yet-unconfirmed prediction of CDM that there are a large abundance of small-scale collapsed halos below the galaxy formation threshold.
Deviations from the CDM predictions at mass scales of $\sim10^7M_{\odot}$, where galaxy formation models have minimal impact on the structure and abundance of dark matter halos, would require physics beyond $\Lambda$CDM with a particle physics model of dark matter.

The fact that individual experiments probe different cosmic epochs allows one to disentangle \textit{ab inito} and \textit{in situ} dark matter physics at scales beyond the galaxy formation threshold.
A consistent deviation of small-scale structure from CDM directly probes the
\emph{ab initio} DM nature and hence such observations are sensitive to the DM thermal history (see Sections \ref{sec:WDM} and \ref{sec:non-therm}), as well as to any nongravitational interactions that DM particles experienced in the early Universe (see Sections \ref{subsec:bar_neut_scat} and \ref{subsec:composite}).
Deviation from CDM originately solely at late times would directly point towards a \textit{in situ} nature of dark matter and can quantify e.g. SIDM (Section~\ref{sec:SIDM}). Hybrid impacts (Section~\ref{sec:hybrid_mods}) can also be constrained.

\section{Stellar and Planet Mass Scale}\label{sec:subsolar}

\subsection{Scale and Objects}

The standard cosmological paradigm with (1) a nearly scale-invariant adiabatic primordial density fluctuations and (2) collisionless cold dark matter (CDM) predicts that dark matter is clustered into halos with a power-law mass spectrum extending from massive galaxy clusters ($10^{15} M_{\odot}$) down to planet-scale masses.\footnote{The absolute lower limit is not known since $N$-body simulations 
suffer from resolution effects at these small scales~\citep{vandenBosch:2017ynq}.} If the dark matter is composed of canonical weakly interacting massive particles (WIMPs) with mass $\sim100$~GeV, the minimum halo mass set by the free-streaming length is $\sim10^{-6} M_{\odot}$ \citep{Green:2003un}.
This simple model has proved adequate to explain observations from large-scale structure, down to the least luminous galaxies that reside in halos of $\sim10^8 M_{\odot}$\citep{Kim:2017iwr, Newton:2018, Jethwa:2018, Nadler:2020prv}. Over the next 5--10 years, observations of ultra-faint galaxies, galaxy--galaxy strong lensing, stellar stream perturbations, and 21cm tomography will all probe dark matter halos that are small enough ($10^6-10^8 M_{\odot}$) to contain few if any stars (see Section~\ref{sec:beyond_galaxy_threshold}) .

Robust identification of dark matter structures well below the threshold of galaxy formation is a qualitatively new observational frontier that is currently mostly unexplored. Measurements in this regime are generically sensitive to a wide range of models that predict damped or enhanced structure --- see the discussion in Sec.~\ref{sec:theory}.
While a suppression of DM structure compared to $\Lambda$CDM on sub-solar scales is well-motivated in the aforementioned scenarios, all existing and proposed probes of these scales cannot reach even the $\Lambda$CDM level of substructure, so they will be difficult to probe. We therefore focus on scenarios with enhanced structure instead. These models offer the exciting opportunity to study the dark sector directly through its gravitational interactions alone. Enhanced structure originates from either an increased initial density contrast leading to an earlier gravitational collapse, or through DM self-interactions that enable the formation of dense clumps, or some combination of the two channels. Models of the former type include cosmological modifications like EMDE (see Sec.~\ref{sec:non_standard_cosmology}) and inflationary production of vector bosons; in this case the characteristic mass scale of DM substructure is connected to cosmological and microphysical parameters like the duration of EMD~\cite{Erickcek:2011us} or the mass of the DM particle~\cite{Graham:2015rva}. Cosmological and model-specific constraints then require that the mass of DM subhalos with enhanced density satisfy $M \lesssim 10^{-3} M_{\odot}$. The characteristic density of these objects is determined by the time of their gravitational collapse which can begin at matter-radiation equality or even earlier (if the initial density contrast is $\gg 1$); in contrast, typical $\Lambda$CDM halos with these masses only form for $z < 30$. Since the matter density scales as $(1+z)^3$, subhalo density can be orders of magnitude larger than in the standard cosmology. In models where DM self-interactions are important, a much broader range of substructure mass and density is possible. For example, atomic dark matter models can produce clumps with $M\gtrsim 10^{-3 }M_\odot$ and densities far larger than $\Lambda$CDM halos (e.g., a dark molecular cooling mechanism can lead to clump densities of $\sim 10^{15}\;\GeV/\mathrm{cm}^3$~\cite{Ghalsasi:2017jna}); similarly, axion scenarios with non-quadratic potentials generate subhalos up to $\sim 10^6$ times denser than $\Lambda$CDM analogues across $10^{-24}< M/M_\odot < 10^9$~\cite{Arvanitaki:2019rax}. Clearly there is a rich portfolio of models that can be tested by probing substructure on sub-solar mass scales.

\subsection{Observational Probes}

Below, we highlight several proposed experimental approaches that could achieve sensitivity to stellar- and planet-mass-scale extended dark matter structures by detecting their gravitational influence on a variety of observables.

\subsubsection{Precision astrometry}

Galactic subhalos can induce detectable proper motions in luminous background sources (\emph{e.g.}, stars and QSOs) due to gravitational lensing\citep{VanTilburg:2018ykj,Mondino:2020rkn, Mishra-Sharma:2020ynk}. 
Several methods have been identified to extract this signal with the ultimate goal of characterizing the lensing subhalo population, \emph{e.g.}, identification of  proper motion outliers due to lensing by compact objects, the use of matched filters to identify extended subhalos from locally correlated proper motions\citep{VanTilburg:2018ykj,Mondino:2020rkn}, and the spectral decomposition of the astrometric velocity and acceleration fields over large regions of the sky~\citep{Mishra-Sharma:2020ynk}. A proof-of-principle analysis using the matched filter approach and data from {\it Gaia} DR2 was recently presented~\citep{Mondino:2020rkn}, demonstrating sensitivity to $\sim 10^8 \Msun$ subhalos extended at the $\sim$\,pc level---much less dense than what established methods relying on photometric microlensing are typically sensitive to. The different proposed methods are complementary and, taken together, will be able to probe a wide variety of scenarios---from NFW subhalos down to $\sim 10^6 \Msun$ to populations of compact objects---when applied to future astrometric datasets such as those from the proposed space-based ESA {\it Theia} mission\citep{Theia:2017}, the upcoming NASA Nancy Grace Roman Space telescope, and radio observations from Square Kilometer Array (SKA).

\subsubsection{Cluster caustics}

Gravitational lensing by galaxy clusters produces caustics in which background objects are magnified by orders of magnitude, and the properties of the images provide a sensitive probe of the mass distribution in the lenses. Observations of background galaxies crossing through such caustics have already been employed to constrain the fraction of DM in compact objects \cite{Oguri:2017ock}, and similar techniques may be used to probe DM substructure. In particular, if the mass distribution of the lens is smooth, images appear in symmetric pairs across critical curves. Thus, asymmetries between these images probe inhomogeneities in the lens, and with detailed modeling, such asymmetries can be related to the abundance of substructure.

Analysis of asymmetries in the caustic-crossing arc SGAS J122651.3+215220 suggests that they are best accounted for by DM subhalos of mass $10^6$--$10^8\Msun$ \cite{Dai:2020rio}. 

Time-domain observations of caustic-transiting stars can be sensitive to DM substructure on solar and sub-solar scales~\cite{Diego:2017drh,Venumadhav:2017pps,Oguri:2017ock}. These stars are highly magnified (reaching absolute of magnifications of $|\mu| \sim \mathcal{O}(10^3-10^4))$, thereby making them extremely sensitive to fluctuations in the lens convergence field (at the level of $|\mu|^{-1}$) and therefore the underlying DM substructure. The macro-caustic due to the cluster lens is expected to be broken up into a network of micro-caustics by the presence of microlenses (stars, black holes, or other compact objects). When a magnified source star is microlensed by such a compact object, the time-variability of its apparent luminosity enables the star to be identified. The event lasts from days to weeks, during which time the source image is stretched by a factor of $\sim |\mu|$ and moves a distance $\sim |\mu| v_{rel} \tau$, where $v_{rel} \sim 1000\;\mathrm{km}/\mathrm{s}$ is the relative velocity between source and lens, and $\tau$ is the duration. This means that the image brightness is sensitive to density fluctuations on length scales from $\sim 4\times 10^4\; \mathrm{AU}$ down to $\sim 5000\; \mathrm{AU}\; (R_\star/100 R_\odot)$, where $R_\star$ is the source radius~\cite{Dai:2019lud}. The maximum size of lensing convergence fluctuations on these scales, $\Delta_\kappa$, is determined by the fraction of DM in subhalos, $f$, and their surface density $\sim M/r_s^2$ relative to the critical density $\Sigma_{cr}$:\footnote{The critical density $\Sigma_{cr}$ is the lens density needed to induce strong lensing.} $\Delta_\kappa \sim \sqrt{f (M/r_s^2)/\Sigma_{cr}}$. This can easily exceed $1/|\mu|$ for interesting examples of DM subhalos. These tiny density fluctuations therefore correspond to $\mathcal{O}(1)$ brightness fluctuations.

Microlensing of highly magnified stars can be used to place constraints on DM substructure both from the observation of the absolute magnitude of the microlensing events, and from the properties of the light curve in an individual event. The former approach has already been used to place bounds on DM fraction in compact objects in Ref.~\cite{Oguri:2017ock}. In Ref.~\cite{Dai:2019lud} it was argued that fluctuations in the microlensing event lightcurve are directly related to the matter power spectrum at small scales, as described above. This is potentially an extremely powerful technique that is sensitive to non-compact structures which are unable to cause strong lensing effects by themselves (as opposed to, for example, black holes).  While the cadence of current observations of extremely-magnified star candidiates~\cite{Kelly:2017fps,Kaurov:2019alr,Chen:2019ncy} are insufficient to constrain this power spectrum, future measurements can access DM substructure with $M \lesssim 10^{-3}M_\odot$ and densities within an order of magnitude of $\Lambda$CDM substructure. 
Sensitivity estimates of this technique in the context of several models (based on Ref.~\cite{Dai:2019lud}) can be found in, e.g., Refs.~\cite{Arvanitaki:2019rax,Blinov:2019jqc,Blinov:2021axd,Xiao:2021ftk}.

The Vera C. Rubin Observatory will detect many more such cluster caustic lens systems. Microlensing events are expected to occur at least once a year for thousands of years~\cite{Venumadhav:2017pps}; therefore continued high-cadence photometric monitoring of these objects can yield unprecedented insight about the DM distribution on sub-solar scales.

\subsubsection{Gravitational microlensing}

Gravitational lensing due to the transit of a compact dark matter object near the line of sight to a source star may give rise to one or several lensed images with an enhanced total apparent luminosity. By convention, a microlensing event is defined by a fractional magnification of $\mu=1.34$. From the non-observation of microlensing events, constraints can be derived on compact dark structures, a technique mostly studied for primordial black holes. 
Recently, such constraints have generalized to extended dark matter structures, such as subhalos, boson stars, and axion miniclusters~\cite{Fairbairn:2017dmf,Croon:2020wpr,Croon:2020ouk}. 
The sensitivity to extended objects depends on their mass profile, with stronger constraints existing for more sharply peaked profiles. 
Current constraints exist for substructures as large as $10^3 \, {\rm R}_\odot$ or about $10$ AU for subhalos of $10^{-2}-10^1 {\rm M}_\odot$~\cite{Croon:2020ouk}, and reach down to $10^{-11} {\rm M}_\odot$ for smaller structures.  CDM subhalos would have radii of $\sim$pc in this range and are thus currently unconstrained through this probe. 

Importantly, the sensitivity of microlensing experiments depends strongly on total observation time and cadence. Longer cadences provide greater sensitivity to more massive objects, and increased observation time increases the overall sensitivity of the survey. Dedicated microlensing studies could also take specific dark matter profiles into account, which typically lead to modified light curves in events.

\subsubsection{Gravitational waves}

Observations of gravitationally lensed distant quasars and galaxies have a long history of probing the invisible dark matter mass throughout the Universe, with high-resolution imaging of galaxy-scale lenses closing in on the $10^6-10^7\Msun$ scale. The recent advent of gravitational wave astronomy has the potential to push down the sensitivity to small dark matter halos by several orders of magnitude \cite{Dai:2018,Oguri:2020}. Indeed, strongly lensed binary black hole (BBH) merger events detectable in the current LIGO band have periods similar to the Schwarzschild timescale of $\sim10^2-10^3\Msun$ compact dark matter clumps. This coincidence of scales leads to a distinctive gravitational wave diffraction pattern that is imprinted on the waveform of the gravitational wave signal. Since the lensing cross section increases significantly with larger source redshift, third generation gravitational wave detectors capable of detecting BBH mergers out to $z_{\rm s} \sim 2-4$ offer the most promising path forward for this technique. Lensed fast radio bursts are expected to be sensitive to low-mass structures via similar diffractive effects \citep{Munoz:2016tmg,Katz:2019qug,Laha:2018zav}.

\subsubsection{Pulsar timing arrays}

Recently, pulsar timing observations have been used to directly measure Galactic accelerations \citep{Phillips:2020xmf,Chakrabarti2021} and the smooth component of the Galactic potential.  In the future, long-term measurements of the time-of-arrivals of light pulses from arrays of pulsars can be used to provide some of the most sensitive probes of substructure at small scales. Dark matter clumps moving around the Earth-pulsar system imprint characteristic shapes in the timing residuals measured by pulsar timing arrays and can be used to search for small-scale structure in the galaxy by their gravitational influence 
alone~\cite{Seto:2007kj,Siegel:2007fz,Baghram:2011is,2012MNRAS.426.1369K,Clark:2015sha,Schutz:2016khr,Dror:2019twh,Ramani:2020hdo, Lee:2020wfn}. 
Clumps passing close by to the Earth or pulsar accelerate the astrophysical body, resulting in a Doppler shift in the pulsar frequency. Additionally, if the clump passes through the line of sight it will cause a Shapiro time delay in the signal. 
This technique is sensitive not only to very compact objects such as black holes, but also to more diffuse objects such as DM minihalos~\cite{Baghram:2011is,Clark:2015sha,Dror:2019twh,Ramani:2020hdo}.  

Standard $\Lambda$CDM sub-halos have been shown to be a challenging target for Pulsar Timing Arrays (PTAs) \cite{Lee:2020wfn}, mostly because the local abundance of these objects is drastically reduced by tidal disruption. 
However, models featuring a population of compact and early-forming sub-halos are expected to be within the reach of near-future PTAs thanks to the reduced impact of tidal effects on these ultra-dense objects.

While the viability of PTAs as probes of DM sub-halos has been established, progress still needs to be done to implement search strategies within the pipelines of PTAs collaborations. Recently a search for sub-halos with $M>10^{-11}M_\odot$ has been implemented by using the analysis code \texttt{enterprise} developed by the North American Nanohertz Observatory for Gravitational Waves (NANOGrav) \cite{Lee:2021zqw}. For these masses, the signal induced by a transiting sub-halos is large enough to be individually resolved, which allows for an easy parametrization of the signal shape. However, PTAs are expected to be sensitive to sub-halos as light as $M\sim10^{-13}M_\odot$. The individual signal induced by these light sub-halos is not large enough to be individually resolved. However, the cumulative signal induced by the entire population could leave a detectable imprint in the timing residuals. The implementation of searches for this signal, which would look like a non-stationary noise in the residuals, is still underway.

\subsubsection{Lensed fast radio bursts}

Fast radio bursts (FRBs) are short and bright electromagnetic emission arriving from cosmological distances (see e.g.,~\cite{Cordes:2019cmq} for a review).
Despite their unknown origins, FRBs are expected to advance our knowledge of dark matter and cosmology over the next decade~\cite{Ravi:2019acz,Munoz:2018mll,Walters:2017afr,Li:2017mek,Abadi:2021ysz}.
Current facilities are detecting thousands of these events (for instance, the Canadian HI Mapping Experiment CHIME~\cite{CHIMEFRB:2021srp}), enough to perform statistical studies.

Of particular interest is the short temporal structure of FRBs, which allows us to study gravitational microlensing on the time domain (rather than the usual spatial one, e.g.~\cite{MACHO:2000qbb}).
As a consequence, FRBs have been proposed as powerful targets to search for DM-induced strong lensing~\cite{Munoz:2016tmg}.
The optical depth for a usual FRB to be lensed by DM, if all of it was composed of point-like lenses (for instance black holes), is $\tau_{\rm lens}\sim 10^{-2}$. As a consequence, a fiducial observation of $10^4$ FRBs (as expected from CHIME) can test any fraction of DM in compact objects down to 1\% of the total DM abundance.
Direct microlensing (i.e., searching for a lensed FRB echo) can detect lenses with masses $M_{\rm lens}\gtrsim 10\,M_\odot$~\cite{Munoz:2016tmg} (see~\cite{Krochek:2021opq,Zhou:2021ndx} for current searches), whereas ``femtolensing" (where the time delay is shorter and fringes appear in the FRB spectra) can test the range $M_{\rm lens}\sim 10^{-4}-10\,M_\odot$~\cite{Katz:2019qug}.
Searching for fuzzier lenses (for instance NFW halos) requires more FRBs, as the optical depth is reduced if the mass of the lens is not entirely contained within its Einstein radius.
As a consequence, more futuristic observatories (beyond CHIME or HIRAX) are likely required to measure NFW halo structure with FRBs, given the CDM prediction.

\subsection{Theory Connections}

For the sub-stellar and sub-Earth halo mass range, the most promising theoretical models to test with current and next-generation facilities are those that predict enhanced central densities relative to CDM halos.
The majority of observational probes at this extremely low mass scale utilize gravitational lensing effects that are strongest for compact mass distributions. 
Already, several of the techniques described here have achieved sensitivity to stellar and planet mass primordial black holes, which can be viewed as a limiting case for the most centrally concentrated halos.
In addition to primordial black holes (discussed in more detail in another Snowmass white paper), other particle physics models to target in the next decade include vector DM produced during inflation, post-inflationary production of QCD axions via the misalignment mechanism, and post-inflationary production of axions via strings (Section \ref{sec:non-therm}), as well as non-standard cosmologies with early matter domination (Section \ref{sec:non_standard_cosmology}).

We advocate for enhanced theoretical, instrumentation, and data analysis efforts to develop experimental techniques to access stellar- and planet-mass-scale extended dark matter structures.
To enable the success of these experiments, it is \emph{essential for the community to support dedicated theory and simulation efforts}, both (1) to model the matter distribution and observables at extremely small scales, and (2) to differentiate between various dark matter and early Universe physics models using detailed and diverse sets of observations.
At this stage, we identify multiple experimental approaches be explored in parallel to better quantify their sensitivity, robustness, and long-term potential, and to check for consistency between signals.

\section{Conclusion}\label{sec:conclusion}

In this white paper, we have presented a wide-ranging set of astrophysical observations that probe the composition and structure of dark matter halos. Among these diverse probes are those that currently provide the strongest evidence of the existence of dark matter to date, and each of them has exciting potential to push boundaries and enable discoveries in the near future. 
\textbf{Importantly, the macroscopic distribution of distribution of dark matter is sensitive to both the fundamental properties of dark matter and its production mechanism in the early Universe.}

We have grouped the different probes in four different mass ranges, from the most massive bound structures of galaxy clusters, down to individual galaxies, pushing below the galaxy formation threshold where we expect halos to be depleted of stars and gas, and down to the stellar and planet mass scale.
The ensemble of astrophysical experiments can probe the entire mass spectrum where we expect imprints of dark matter in our Universe.
The key opportunity lies in the combination of multiple different measurements sensitive to different mass scales. A holistic approach of a combined probe analysis and modeling enables to pinpoint a potential non-gravitational signature of dark matter and directly link the macroscopic emerging cosmological structure with the microscopic particle nature of dark matter.

Given that the different individual experiments are also sensitive to different epochs of the Universe, the originating microphysical cause of the effect can be disentangled between \textit{ab initio} where early-Universe DM behavior modifies the initial conditions for structure formation and \textit{in situ} where a relatively high density of DM in the present-day Universe allows certain DM behaviors to become manifest.
Overall, the astrophysical dark matter experiment probes an enormous range of dark matter models of which only few specific examples have been mentioned in this white paper. There are more opportunities to develop the phenomenology in terms of the early Universe as well as late-time evolution.

Complementary to the core message of this white paper linking astrophysical observations with dark matter microphysics, the community relies on state-of-the-art facilities and instrumentation together with accurate and precise simulation predictions. These topics are addressed in two accompanying Snowmass white papers.

\bibliographystyle{JHEP.bst}
\bibliography{main.bib}

\end{document}